# Theoretical Analysis for the CommSense Measurement System

Sandip Jana, *Graduate Student Member, IEEE*
Amit Kumar Mishra, *Senior Member, IEEE*
Mohammed Zafar Ali Khan, *Senior Member, IEEE*

***Abstract*—Modern wireless networks can blur the boundary between communication and sensing by treating OFDM transmissions as calibrated "measurement signals" that serve both high-throughput data delivery and environmental monitoring. In this work, we present a rigorous theoretical analysis of the Communication based Sensing (CommSense) instrument, which integrates lightweight PCA-based detector modules directly into standard OFDM receivers. This enables real-time, device-free measurement of passive scatterers (e.g. drones, vehicles) using only existing downlink reference symbols, no additional transmitters or spectrum allocation are required. Starting from a realistic three-link Rician channel model (direct Tx→Rx, cascaded Tx→Scatterer, and Scatterer→Rx), we derive and compare four detector architectures: the full-dimensional Likelihood Ratio Test (Full LRT), PCA-based LRT (arising naturally from the statistical framework), and the CommSense variants: PCA+SVM with linear and RBF kernels. By projecting the $N$-dimensional CSI measurement vector onto a $P \ll N$ principal subspace, the instrument reduces inference latency by an order of magnitude relative to the Full LRT, while maintaining optimal error performance: empirical error rates track the Bhattacharyya bound and the Area Under the ROC Curve reaches $\approx 1$ for $P \approx 10$. In very high-dimensional scenarios ($N = 1024$), PCA + SVM classifiers deliver AUC$\gtrsim 0.60$ at $-10$ dB and exceed 0.90 by 0 dB, where Full LRT succumbs to numerical overflow. Our results further reveal that LRT-based measurement approaches are sensitive to parameter estimation errors, whereas the PCA-driven SVM classifiers exhibit strong resilience. Altogether, this study demonstrates that the CommSense measurement system combining PCA for subspace compression with lightweight SVM inference constitutes a fast, accurate, and robust instrumentation solution for scatterer sensing, paving the way for integrated sensing and communication (ISAC) in next generation of wireless systems.**

## I. Introduction

ISAC, also known as Joint Communication and Sensing (JCAS), is evolving into a fundamental measurement technique within modern wireless systems, effectively instrumenting the radio as both a data link and an environmental probe. In this instrumentation paradigm, standard OFDM waveforms act as calibrated "test signals" that illuminate the propagation environment, enabling quantitative measurement of physical phenomena: such as vehicle movement, drone presence, and other dynamic events in real time. By co-designing the signal generation, channel-sounding front end, and post-processing algorithms, ISAC repurposes conventional communication hardware into a unified sensing and data-acquisition system, delivering seamless connectivity alongside continuous situational awareness.

This work was supported by University Grants Commission (UGC) and partly by project VAJRA, Department of Science and Technology (DST), Govt. of India.

Sandip Jana and Mohammed Zafar Ali Khan is with Department of Electrical Engineering, Indian Institute of Technology, Hyderabad, India.

Amit Kumar Mishra is with National Spectrum Centre, Aberystwyth University, Aberystwyth, U.K. He is also a visiting professor at Division of Computer Science, University West, Sweden

### A. Motivation and Evolution of CommSense System

CommSense measurement system represents a task-specific offshoot of the broader ISAC paradigm, but with a sharper focus on environmental inference using existing wireless signals. As a new type of measurement instrument, it treats ambient communication waveforms: cellular downlinks, Wi-Fi beacons, and other signals of opportunity as non-intrusive "illumination sources" for sensing, rather than dedicating additional transmit hardware. While ISAC traditionally emphasizes joint waveform design for both data transmission and radar functionality, CommSense measurement system leverages in-situ channel models and the Application Specific INstrumentation (ASIN) framework [1] to optimize purely for sensing performance. As depicted in Fig.1, the ASIN philosophy is the cornerstone of our instrument: by pairing low-resolution sensors with early, task-specific feature extraction, we achieve rapid detection without the need for high-fidelity measurements. This specialization enables efficient on-edge operation; minimizing hardware complexity, data volume, and energy consumption for our target application.

To date, published works have demonstrated proof-of-concept and experimental results for communication-based sensing, but until now there has been little to no thorough theoretical analysis of this measurement technique. In particular, naïve application of a full-dimensional Likelihood Ratio Test (Full LRT) across all $N$ OFDM subcarriers incurs $\mathcal{O}(N^3) + \mathcal{O}(N^2)$ complexity and suffers severe numerical instability in large-scale systems ( [2], Sec. III-G), rendering real-time inference impractical.

Our work bridges this gap by presenting the comprehensive theoretical framework for the CommSense measurement system. Rather than focusing on detection alone, we derive the underlying theory that gives rise to several detectors: Full LRT and PCA+LRT being direct consequences of the statistical model, while the CommSense variants (PCA+SVM classifiers) perform inference without requiring accurate knowledge of link statistics. Hence the CommSense variants are robust against inaccurate estimate of the statistical parameters. Specifically, we project the high-dimensional CSI data onto a low-rank subspace via Principal Component Analysis (PCA), then apply lightweight discriminative classifiers (e.g. linear or RBF-kernel SVM) on only $P \ll N$ components.

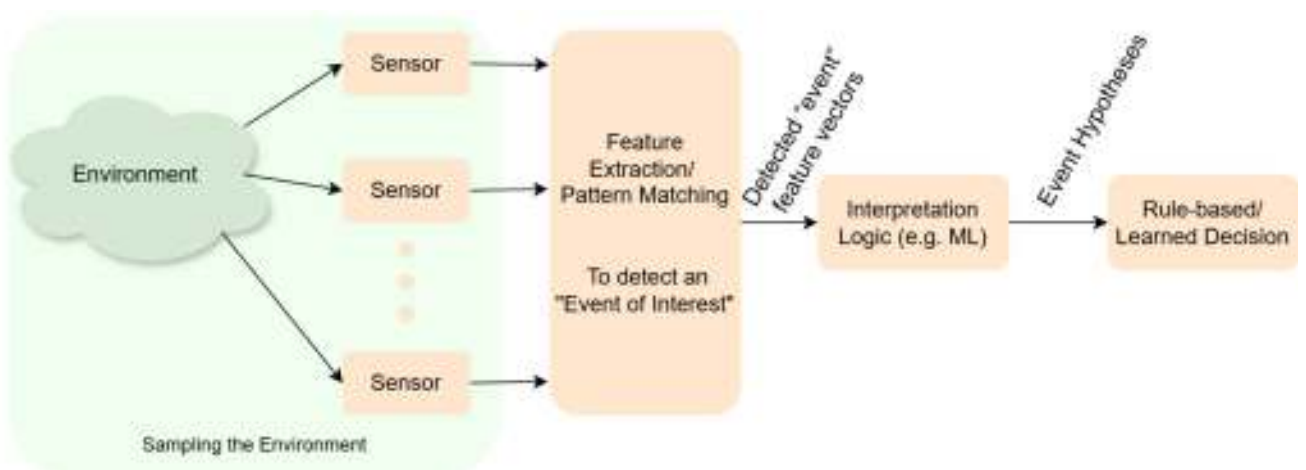

Fig. 1. Block diagram of the ASIN framework: a bio-inspired, task-driven sensing paradigm that prioritizes "detect-first, identify-later" processing with lightweight ML models [1]

Key Benefits of CommSense System:

- **Zero Footprint Emission:** No extra transmitters: CommSense piggybacks on ubiquitous communication signals.
- **Cost Effective Deployment:** Leverages off-the-shelf SDR/UE hardware and existing network infrastructure.
- **Regulatory Simplicity:** Avoids spectrum-licensing and transmission-power approvals by using only the downlinks.
- **Seamless Integration:** Runs side-by-side with standard communications pipeline, yielding dual use of the same RF front end.
- **Lightweight AI/ML:** Employs low-dimensional PCA features and compact SVM/LRT classifiers for real-time, high-accuracy detection under challenging conditions.

### B. State of the Art

Passive sensing with communication waveforms has attracted significant interest for device-free detection and localization. Early approaches relied on simple, low-dimensional features extracted from WiFi or OFDM channel state information (CSI). For example, authors in [3] used received signal strength (RSS) for indoor tracking and navigation. [4] proposes a room-level, cost-effective Bluetooth-based tracking method that avoids fingerprint calibration, achieving a 94.2% hit rate with minimal 3.6%false alarms. Authors in [5] propose an ML-based indoor/outdoor (IO) sensing and positioning framework that uses a random forest to filter fingerprints and a WKNN-enhanced Cell ID algorithm on denoised measurement reports, achieving notably lower positioning errors than conventional fingerprinting methods. [6] proposes a cost-effective OTA ISAC testbed that reuses a standard channel emulator to generate deceptive echoes and support communication testing for joint evaluation. Experimental delay/Doppler emulations and a vehicle-mounted mm-Wave radar trial confirm its ability to replicate realistic propagation scenarios for ISAC Base Station testing. In [7], the authors propose a JCAS solution for smart agriculture, combining an ultra-compact soil moisture sensor (UCSMS) for high-precision measurements with a pattern-reconfigurable antenna (PRA) to relay data efficiently to the base station. [8] presents sensing humans among passive radio frequency (SHAPR), capturing human-sensitive PRF spectra by using a SDR and applying five ML algorithms for classification and regression across various environments. [9] implements a real-time wireless synchronization scheme on an FPGA platform, which provides highly precise time-frequency synchronization, vital for networked ISAC systems. [10] enhances WiFi-based indoor localization by reducing radio mapping efforts through Gaussian process regression for RSS prediction, improving fingerprint accuracy with compound kernels, and determining precise locations using a weighted Similarity K-Nearest Neighbor algorithm. Experimental results show significant localization error reduction without extra calibration or infrastructure. While these methods are computationally light, they degrade rapidly at low SNR since they cannot fully exploit the spatial and frequency diversity inherent in OFDM CSI.

To address these limitations, subspace based techniques have been developed. [11] perform principal component analysis (PCA) of the CSI covariance matrix to select the most *sensing sensitive* directions for occupancy detection, achieving improved robustness at moderate complexity. The authors propose a unique LTE based non-intrusive, low-cost, passive indoor occupancy estimation solution that first determines whether the indoor environment is empty or not and, if occupied, estimates the number of people present. In [12], authors implement a similar principal component filtering to detect static scatterers and drones, and achieving up to 99% accuracy across various environments and distances without needing dedicated transmitters.

Unlike many ISAC efforts, CommSense has been under continuous development in our group for over a decade: evolving from early GSM based demos [13] to LTE crowd size estimation experiments [14], [15], 5G NR feasibility studies [16], and even 60GHz mmWave trials [17]. These campaigns have validated CommSense's versatility: for instance, behind-the-wall person and weapon detection with accuracies of 77.5% and 95.2% using 577 µs GSM frames [13], outdoor crowd estimation via LTE echoes during the COVID-19 post-lockdown phase [14]. Together, these results attest to CommSense's practical robustness and its promise for real-time sensing in 6G and beyond.

In summary, while subspace selection and ISAC frameworks recognize the low-rank nature of the scatterer signature, there remains a gap in systematically evaluating *dimensionality reduced* detectors that combine statistical optimality with computational efficiency. Our work closes this gap by developing and benchmarking PCA+LRT and PCA+SVM schemes, demonstrating that only $P \ll N$ dimensions are sufficient to attain near Bayes-optimal detection at a fraction of the cost of full-dimensional methods.

### C. Main Contributions and Structure of this paper

The main contributions of this paper are the following:

- **Realistic CommSense Model:** For theoretical analysis, we develop a three link Rician channel model for scat-

terer sensing (Tx$\rightarrow$Rx direct, Tx$\rightarrow$Scatter, Scatter$\rightarrow$Rx cascaded along with scatterer RCS).

- **PCA-Accelerated Detectors:** We give the rationale behind choosing reduced-complexity detectors: PCA+LRT and PCA+SVM (linear and RBF kernels), which project the high dimensional CSI onto a low-rank subspace before classification.
- **Computational Efficiency:** We demonstrate that PCA based methods achieve an order of magnitude faster inference than the full-dimensional LRT.
- **Robustness in High Dimensions:** Through extensive simulations for $N = 256$ and $N = 1024$ subcarriers over SNRs from $-10\,\text{dB}$ to $15\,\text{dB}$, we show that PCA+SVM maintains near optimal ROC AUC even when the full LRT fails numerically in very large dimensions.
- **Theoretical Validation:** We derive and compute the Bhattacharyya bound before and after PCA, and confirm that empirical error rates closely follow the predicted limits, highlighting that PCA retains the discriminating characteristics in the reduced subspace.
- **Practical Design Guidelines:** We identify an "elbow" in the PCA spectrum at $P \approx 10$, beyond which further components are noise-dominated, and provide concrete recommendations for choosing $P$ to balance speed and accuracy in real-time scatterer sensing.

The remainder of this paper is structured as follows: In Chapter II, we will detail the CommSense architecture, illustrating how ambient OFDM signals from the physical environment are captured and transformed into digital channel state information for sensing. Chapter III lays the theoretical groundwork, deriving the complex-Gaussian models, the full-dimensional Likelihood-Ratio Test, and the Bhattacharyya bound for scatterer detection. Chapter IV presents a series of simulations that compare full-dimensional inference against PCA-based methods (PCA+LRT and PCA+SVM), demonstrating how a low-rank subspace preserves the essential scatterer signature while slashing computational cost. Finally, Chapter V summarizes our findings, highlights practical design guidelines for real-time OFDM sensing, and outlines avenues for future research.

## II. CommSense Cyber-Physical Architecture

Fig. 2 depicts the end-to-end CommSense architecture, which bridges the *Physical World* of RF propagation and scatterer interactions with a *Digital Sensing Pipeline* running alongside standard LTE communications:

- **Physical Layer Illumination:** An LTE eNodeB transmits OFDM signal $s(t)$, which comprises the modulated symbols $S_n$ over $N$ subcarriers. The composite channel $g(t)$ comprises a line of sight (LOS) tap and potentially multiple scatterer taps (e.g. from buildings, drones, vehicles etc.), each with its own delay $\tau$, Doppler $f_D$, and Rician $K$-factor.
- **Capture via SDR UE:** A software defined radio (SDR) based LTE UE receives the downlink, performs coarse synchronization and FFT to obtain the channel frequency response estimates at pilot subcarriers:
$$\hat{\mathbf{G}} = [\hat{G}_0, \hat{G}_1, \ldots, \hat{G}_{N-1}]^T,$$
- **Communication Pipeline:** In parallel, the UE runs standard symbol equalization and demodulation, computing metrics such as bit error rate (BER) and throughput.
- **Sensing Pipeline:**
    1) *CSI Extraction & Database Generation:* To detect an event of interest, gather $\hat{\mathbf{G}}$ under the null hypothesis $\mathcal{H}_0$ and alternate hypothesis $\mathcal{H}_1$ (i.e. event has occured) over $M$ OFDM symbols.
    2) *Principal Feature Extraction:* Perform PCA on the standardized CSI matrix, obtaining the top $P \ll N)$ singular vectors $\mathbf{U}_P$, and form reduced dimension features.
    3) *Detector Training:* Use extracted features to train
        - PCA+LRT: Gaussian LRT in $\mathbb{C}^P$,
        - PCA+SVM with Linear or Radial Basis Function (RBF) kernels: discriminative classification.
    4) *Event Detection & Evaluation:* Apply the trained classifier to incoming CSI estimates, outputting detection decisions. Compute performance metrics (detection accuracy, Receiver Operating Characteristic (ROC) curve, etc) in real time.

This dual pipeline design allows a single OFDM receiver to simultaneously sustain communications and deliver robust situational awareness via low rank subspace sensing.

## III. Background Theory

OFDM is the modulation of choice in 3GPP standards for both LTE [18] and 5G NR [19] due to its:

- **Multipath Robustness:** The cyclic-prefix OFDM structure effectively mitigates inter-symbol interference in frequency selective channels.
- **Flexible Resource Allocation:** Subcarrier level granularity allows adaptive bit-loading and dynamic bandwidth partitioning.
- **Rich CSI Availability:** Pilot subcarriers embedded across the OFDM grid provide high resolution channel estimates without any additional signaling.

These properties make OFDM an ideal choice for CommSense, enabling passive, real-time environmental sensing. Hence, in this section, we will develop a comprehensive mathematical framework for scatterer detection using OFDM signals.

### *A. Transmitted OFDM Signal*

An Orthogonal Frequency Division Multiplexing (OFDM) symbol consists of $N$ subcarriers, each modulated with a complex I-Q symbol $S_n \in \mathbb{C}$, where $n = 0, 1, \ldots, N-1$. The time-domain OFDM symbol $s(t)$ is obtained by applying the Inverse Fast Fourier Transform (IFFT):

$$s(t) = \sum_{n=0}^{N-1} S_n e^{j2\pi n \Delta f t}, \quad 0 \leq t \leq T_{\text{sym}}, \tag{1}$$

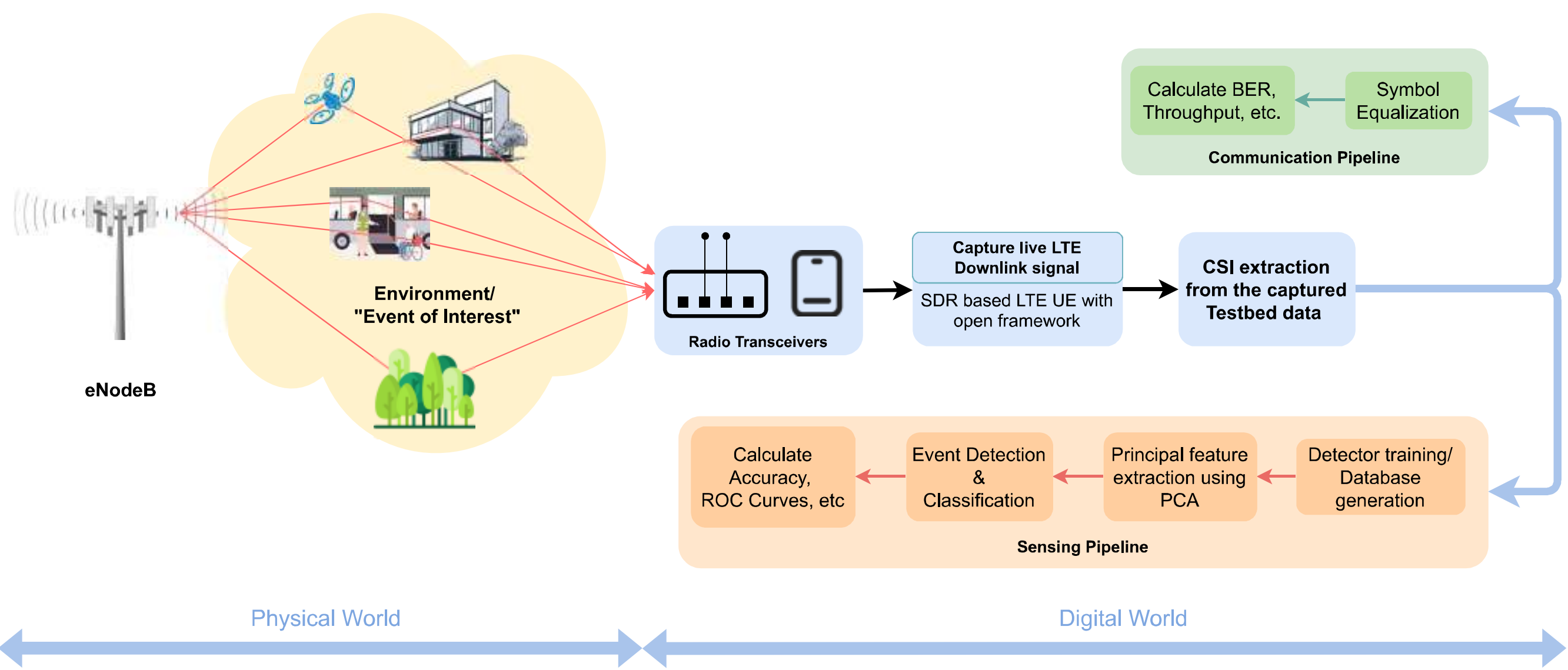

Fig. 2. **CommSense Cyber-Physical Architecture: From RF Scattering to Event Detection.**The left side shows the *Physical World*: an OFDM base station (eNodeB) illuminates an environment containing static clutter and moving scatterers (e.g., vehicles, drones, pedestrians). The multipath returns are captured by a software defined LTE UE. The right side shows the *Digital World*, split into two parallel pipelines: *(top)* the standard *Communication Pipeline* (symbol synchronization, equalization, BER/throughput measurements), and *(bottom)* the *Sensing Pipeline*, which (i) extracts CSI estimates from each OFDM symbol, (ii) builds a training database under hypotheses $\mathcal{H}_0, \mathcal{H}_1$, (iii) computes principal features via PCA, (iv) trains classifiers (LRT or SVM) on the projected data, and (v) performs real-time event detection and reports detection accuracy, ROC curves, etc.

where:

- $\Delta f = \dfrac{1}{T_{\text{sym}}}$ is the subcarrier spacing.
- $T_{\text{sym}}$ is the OFDM symbol duration.

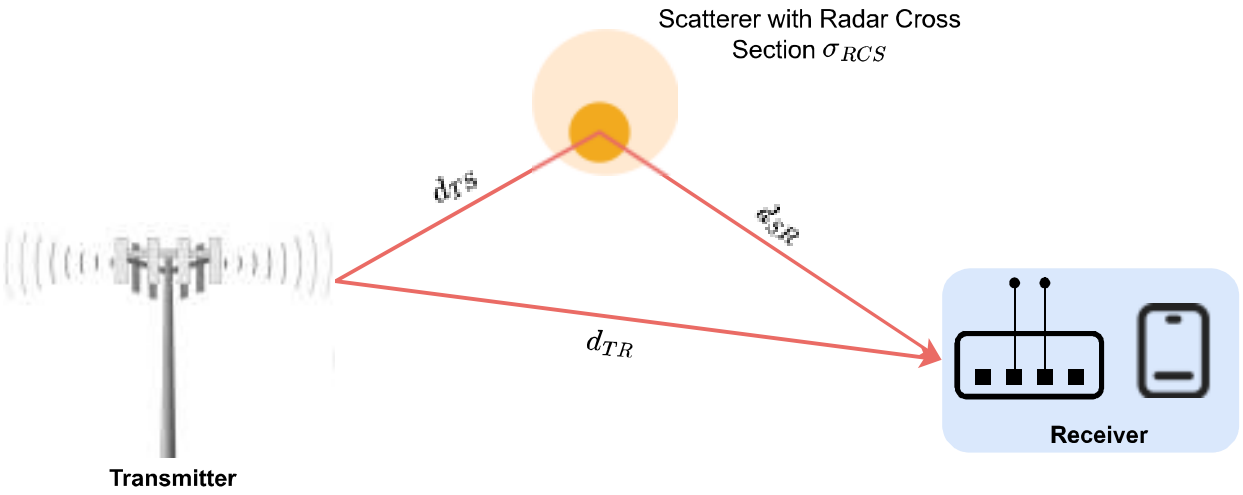

Fig. 3. Transmitter, Receiver, and Scatterer configuration

## B. Wireless Channel

The wireless channel can be modeled as a combination of large-scale and small-scale effects.

*1) Friis path loss:* This model calculates the received power under ideal free-space conditions. In a simple point-to-point communication context, assuming a unobstructed line-of-sight (LOS) between transmitter and receiver. The Friis equation is given by:

$$P_{\text{R}} = P_{\text{T}}\, G_{\text{T}}\, G_{\text{R}} \left(\frac{\lambda}{4\pi d_{\text{TR}}}\right)^2$$
$$\implies A_{\text{R}} = \sqrt{P_{\text{T}}\, G_{\text{T}}\, G_{\text{R}}} \left(\frac{\lambda}{4\pi d_{\text{TR}}}\right) \tag{2}$$

where:

- $P_{\text{R}}$ is the received power,
- $P_{\text{T}}$ is the transmitted power,
- $G_{\text{T}}$ is the transmitter antenna gain,
- $G_{\text{R}}$ is the receiver antenna gain,
- $\lambda$ is the wavelength,
- $d_{\text{TR}}$ is the distance between the transmitting and receiving antennas,
- $A_{\text{R}}$ is the amplitude of the received signal.

The Friis equation can be extended to scenarios where a target with radar cross section (RCS) $\sigma_{RCS}$ scatters the transmitted signal. Under free-space conditions and assuming LOS from transmitter to the scatterer and from the scatterer to the receiver, the received power is given by,

$$P_{\text{R}} = P_{\text{T}}\, G_{\text{T}}\, G_{\text{R}}\, \sigma_{RCS} \left(\frac{\lambda^2}{(4\pi)^3 d_{\text{TS}}^2 d_{\text{SR}}^2}\right)$$
$$\implies A_{\text{R}} = \sqrt{P_{\text{T}}\, G_{\text{T}}\, G_{\text{R}}\, \sigma_{RCS}} \left(\frac{\lambda}{(4\pi)^{3/2} d_{\text{TS}} d_{\text{SR}}}\right) \tag{3}$$

where:

- $d_{\text{TS}}, d_{\text{SR}}$ are the distance between transmitter-scatterer and scatterer-receiver paths.

*2) Small-Scale Fading:* It describes rapid fluctuations in the received signal's amplitude and phase, typically caused by multipath propagation. The channel impulse response (CIR) can be modeled as as a tapped delay line with $L$ taps:

$$g(t) = \sum_{l=1}^{L} g_l\, \delta(t - \tau_l), \tag{4}$$

where $g_l$ and $\tau_l$ are the gain and delay of $l^{th}$ tap and each tap is represented as:

$$g_l = \alpha_l\, e^{j\phi_l},$$

TABLE I
LIST OF VARIABLES AND NOTATIONS

| Variables | Definition |
| --- | --- |
| $f_c$ | Center frequency |
| $\lambda$ | Wavelength |
| N | Total number of subcarriers |
| M | Total number of symbols |
| $T_{\text{sym}}$ | OFDM symbol duration |
| $\Delta_f$ | Subcarrier spacing |
| $P_{\text{R}}$ | Received power |
| $P_{\text{T}}$ | Transmitted power |
| $G_{\text{T}}$ | Transmitter antenna gain |
| $G_{\text{R}}$ | Receiver antenna gain |
| $d_{\text{TR}}, d_{\text{TS}}, d_{\text{SR}}$ | Distance between the Tx$\rightarrow$Rx, Tx$\rightarrow$Scatterer, Scatterer$\rightarrow$Rx links respectively |
| $A_{\text{R}}$ | Amplitude of the received signal |
| $\sigma_{\text{RCS}}$ | Radar Cross Section (RCS) of the scatterer |
| $L$ | Number of channel taps |
| $(g_l, \tau_l)$ | Gain and Delay of the $l^{th}$ tap respectively |
| $K$ | Parameter for Rician distribution |
| $P_{\text{LOS}}, P_{\text{NLOS}}$ | Power in the LOS and NLOS components |
| $g_{\text{LOS}}(t), g_{\text{NLOS}}(t)$ | Channel gain of LOS and NLOS components |
| $\tau_0$ | LOS propagation delay |
| $f_D$ | Doppler frequency, |
| $\theta_{\text{offset}}$ | Initial phase offset. |
| $T_s$ | Sampling Duration |
| $\mathbf{F}$ | DFT Matrix |
| $\mathbf{g}$ | $= [g[0], g[1], \ldots, g[N-1]]^T$, in discrete time domain |
| $\boldsymbol{\mu}_g$ | $= [\mu_g[0], \mu_g[1], \ldots, \mu_g[N-1]]^T$, Mean vector |
| $\boldsymbol{\Sigma}_g$ | Covariance matrix for $\mathbf{g}$ |
| $\mathbf{G}$ | $= \mathbf{F}\,\mathbf{g}$ |
| $\boldsymbol{\mu}_{\mathbf{G}}$ | $= \mathbf{F}\,\boldsymbol{\mu}_g$ |
| $\boldsymbol{\Sigma}_{\mathbf{G}}$ | $= \mathbf{F}\,\boldsymbol{\Sigma}_g\,\mathbf{F}^H$ |
| $(\boldsymbol{\mu}_{\mathcal{H}_i}, \Sigma_{\mathcal{H}_i})$ | Mean and Covariance for $i^{th}$ Hypothesis |

with $\alpha_l$ as the amplitude and $\phi_l$ as the phase.

In a non-line-of-sight (NLOS) scenario, the real and imaginary parts of $g_l$ is modeled as $\mathcal{N}(0, \sigma^2_{\text{NLOS}})$. Thus, the envelope of the received signal follows the Rayleigh distribution.

*3) Combination of LOS and NLOS:* This type of channel is modeled as Rician fading channel where the Rician $K$-factor measures the ratio of the power in the deterministic (LOS) component to the power in the diffuse (NLOS) component. Mathematically,

$$K = \frac{P_{\text{LOS}}}{P_{\text{NLOS}}}. \tag{5}$$

For convenience, we will normalize the total average power of the channel to 1. i.e. $P_{\text{LOS}} + P_{\text{NLOS}} = 1$. Therefore,

$$P_{\text{LOS}} = \frac{K}{K+1}, \quad P_{\text{NLOS}} = \frac{1}{K+1}.$$

Thus the Rician channel becomes,

$$g(t) = \sqrt{\frac{K}{K+1}}\, g_{\text{LOS}}(t) + \sqrt{\frac{1}{K+1}}\, g_{\text{NLOS}}(t) \tag{6}$$

*a) LOS Component:*

$$g_{\text{LOS}}(t) = A\, e^{j\,\theta(t)} \tag{7}$$

where, $A$ is the amplitude of the received signal derived in (2) and (3), and $\theta(t)$ captures any phase rotation.

$$\theta(t) = -2\pi f_c \tau_0 + 2\pi f_D t + \theta_{\text{offset}}$$

where:

- $\tau_0$ is the LOS propagation delay,
- $f_D$ is the Doppler frequency,
- $\theta_{\text{offset}}$ is the initial phase offset.

*b) NLOS Component:* As discussed in the Sec.III-B2,

$$\begin{aligned} \Re(g_{\text{NLOS}}(t)), \Im(g_{\text{NLOS}}(t)) &\sim \mathcal{N}\Big(0, \sigma^2_{\text{NLOS}}\Big) \\ \implies g_{\text{NLOS}}(t) &\sim \mathcal{CN}\Big(0, 2\sigma^2_{\text{NLOS}}\Big) \end{aligned} \tag{8}$$

Now putting (7) and (8) together and ensuring overall Rician channel $g(t)$ in (6) has unit power, we define $K$ as,

$$K = \frac{A^2}{2\sigma^2_{NLOS}} \tag{9}$$

$$g_{LOS}(t) = \frac{g_{LOS}(t)}{|g_{LOS}(t)|} = e^{j\theta(t)} \tag{10}$$

$$g_{NLOS}(t) \sim \mathcal{CN}(0, 1) \tag{11}$$

$$\implies g(t) \sim \mathcal{CN}\Bigg(\sqrt{\frac{K}{K+1}} e^{j\,\theta(t)}, \frac{1}{K+1}\Bigg) \tag{12}$$

In practice, we will work with discrete-time samples $g(n)$, which is simply the continuous-time $g(t)$ evaluated at $t = nT_s$, where $T_s$ is the sampling interval.

$$\begin{aligned} g[n] &\sim \mathcal{CN}\Bigg(\sqrt{\frac{K}{K+1}} e^{j\,\theta[n]}, \frac{1}{K+1}\Bigg) \\ &\sim \mathcal{CN}\big(\mu_g[n], \sigma_g^2\big) \\ \theta[n] &= \theta[nT_s] = -2\pi f_c \tau_0 + 2\pi f_D (nT_s) + \theta_{\text{offset}} \end{aligned} \tag{13}$$

Now consider we have $N$ time samples and assuming i.i.d. in time:

$$\begin{aligned} \mathbf{g} &= [g[0], g[1], \ldots, g[N-1]]^T \\ &\sim \mathcal{CN}\big(\boldsymbol{\mu}_g, \boldsymbol{\Sigma}_g\big) \end{aligned} \tag{14}$$

$$\boldsymbol{\mu}_g = [\mu_g[0], \mu_g[1], \ldots, \mu_g[N-1]]^T, \;\; \boldsymbol{\Sigma}_g = \sigma_g^2 \mathbf{I}_N \tag{15}$$

Let $\mathbf{G}$ be the N-point DFT of $\mathbf{g}$, in matrix form, we can write $\mathbf{G} = \mathbf{F}\,\mathbf{g}$
where $\mathbf{F}$ is the $N \times N$ DFT matrix with entries

$$\mathbf{F}_{k,n} = e^{-j\frac{2\pi kn}{N}}, \quad k, n = 0, 1, \ldots, N-1.$$

Then,

$$\begin{aligned} \mathbf{G} &\sim \mathcal{CN}\big(\mathbf{F}\,\boldsymbol{\mu}_g,\; \mathbf{F}\,\boldsymbol{\Sigma}_g\,\mathbf{F}^H\big) \\ &\sim \mathcal{CN}\big(\boldsymbol{\mu}_{\mathbf{G}},\; \boldsymbol{\Sigma}_{\mathbf{G}}\big) \end{aligned}$$

$$\boldsymbol{\mu}_{\mathbf{G}} = \mathbf{F}\,\boldsymbol{\mu}_g \;=\; \begin{bmatrix} \sum_{n=0}^{N-1} \mu_g[n]\, e^{-j\,\frac{2\pi}{N}\,0\,n} \\ \sum_{n=0}^{N-1} \mu_g[n]\, e^{-j\,\frac{2\pi}{N}\,1\,n} \\ \vdots \\ \sum_{n=0}^{N-1} \mu_g[n]\, e^{-j\,\frac{2\pi}{N}\,(N-1)\,n} \end{bmatrix} = \begin{bmatrix} \mu_G[0] \\ \mu_G[1] \\ \vdots \\ \mu_G[N-1] \end{bmatrix}$$

$$\text{where} \quad \mu_{\mathbf{G}}[k] \;=\; \sum_{n=0}^{N-1} \sqrt{\frac{K}{K+1}}\, e^{j\,\theta[n]}\, e^{-j\,\frac{2\pi}{N}\,k\,n} \tag{16}$$

For the unnormalized DFT matrix $\mathbf{F}$, $\mathbf{F}\,\mathbf{F}^H = N\,\mathbf{I}_N$.

$$\begin{aligned} \boldsymbol{\Sigma}_{\mathbf{G}} &= \mathbf{F}\,\boldsymbol{\Sigma}_g\,\mathbf{F}^H \qquad (17) \\ &= \sigma_g^2\,\mathbf{F}\,\mathbf{I}_N\,\mathbf{F}^H, \;\; \text{Using (15)} \\ &= \sigma_g^2\,(\mathbf{F}\,\mathbf{F}^H) = \sigma_g^2\,N\,\mathbf{I}_N = \frac{N}{K+1}\,\mathbf{I}_N \qquad (18) \end{aligned}$$

### C. Received Signal

The received signal $r(t)$ is affected by the wireless channel and noise, under two hypotheses:

*1) **Null Hypothesis** ($\mathcal{H}_0$):* Ambient environment (no scatterer is present).

$$r(t) = g_{TR}(t) \circledast s(t) + \zeta_0(t) \tag{19}$$

where $\circledast$ is the circular convolution operator. After applying the N-point Discrete Fourier Transform (DFT),

$$\begin{aligned} R_{\mathcal{H}_0} &= G_{\text{TR}}\,S + Z_0 \\ &= G_{\mathcal{H}_0}\,S + Z_0 \end{aligned} \tag{20}$$

*2) **Alternative Hypothesis** ($\mathcal{H}_1$):* Environment with the scatterer present.

$$r(t) = g_{TR}(t) \circledast s(t) + g_{TS}(t) \circledast g_{SR}(t) \circledast s(t) + \zeta_1(t) \tag{21}$$

After taking the DFT,

$$\begin{aligned} R_{\mathcal{H}_1} &= G_{\text{TR}}\,S + G_{\text{TS}}\,G_{\text{SR}}\,S + Z_1 \\ &= (G_{\text{TR}} + G_{\text{TS}}\,G_{\text{SR}})\,S + Z_1 \\ &= (G_{\text{TR}} + G_{\text{cascaded}})\,S + Z_1 \\ &= G_{\mathcal{H}_1}\,S + Z_1 \end{aligned} \tag{22}$$

### D. Link Analysis

From the configuration given in Fig.3 and the distribution given in (13),

*1) Tx-Rx (TR) link:*

$$\begin{aligned} \mathrm{g}_{TR}[n] \;&\sim\; \mathcal{CN}\Big(\sqrt{\tfrac{K_{TR}}{K_{TR}+1}}\;e^{j\,\theta_{TR}[n]},\; \tfrac{1}{K_{TR}+1}\Big) \\ &\sim\; \mathcal{CN}\big(\mu_{\mathrm{g}_{TR}}[n],\; \sigma^2_{\mathrm{g}_{TR}}\big) \end{aligned} \tag{23}$$

After taking N-pt DFT, similar to (16) and (18)

$$\mathbf{G}_{TR} = \mathbf{F}\,\mathbf{g}_{TR} \;\sim\; \mathcal{CN}\big(\boldsymbol{\mu}_{\mathbf{G}_{TR}},\; \boldsymbol{\Sigma}_{\mathbf{G}_{TR}}\big) \tag{24}$$

$$\text{where} \quad \boldsymbol{\mu}_{\mathbf{G}_{TR}} = [\mu_{G_{TR}}[0],\, \mu_{G_{TR}}[1],\, \ldots,\, \mu_{G_{TR}}[N-1]]^T$$

$$\mu_{\mathbf{G}_{TR}}[k] \;=\; \sum_{n=0}^{N-1} \sqrt{\frac{K_{TR}}{K_{TR}+1}}\, e^{j\,\theta_{TR}(n)}\, e^{-j\,\frac{2\pi}{N}\,k\,n} \tag{25}$$

$$\boldsymbol{\Sigma}_{\mathbf{G}_{TR}} = \frac{N}{K_{TR}+1}\,\mathbf{I}_N \tag{26}$$

*2) Tx-Scatterer (TS) link:* Proceeding in the similar way as described above in Sec.III-D1,

$$\begin{aligned} \mathrm{g}_{TS}[n] \;&\sim\; \mathcal{CN}\Big(\sqrt{\tfrac{K_{TS}}{K_{TS}+1}}\;e^{j\,\theta_{TS}[n]},\; \tfrac{1}{K_{TS}+1}\Big) \\ &\sim\; \mathcal{CN}\big(\mu_{\mathrm{g}_{TS}}[n],\; \sigma^2_{\mathrm{g}_{TS}}\big) \end{aligned} \tag{27}$$

$$\mathbf{G}_{TS} = \mathbf{F}\,\mathbf{g}_{TS} \;\sim\; \mathcal{CN}\big(\boldsymbol{\mu}_{\mathbf{G}_{TS}},\; \boldsymbol{\Sigma}_{\mathbf{G}_{TS}}\big) \tag{28}$$

$$\boldsymbol{\mu}_{\mathbf{G}_{TS}} = [\mu_{G_{TS}}[0],\, \mu_{G_{TS}}[1],\, \ldots,\, \mu_{G_{TS}}[N-1]]^T$$

$$\mu_{\mathbf{G}_{TS}}[k] \;=\; \sum_{n=0}^{N-1} \sqrt{\frac{K_{TS}}{K_{TS}+1}}\, e^{j\,\theta_{TS}(n)}\, e^{-j\,\frac{2\pi}{N}\,k\,n}$$

$$\boldsymbol{\Sigma}_{\mathbf{G}_{TS}} = \frac{N}{K_{TS}+1}\,\mathbf{I}_N$$

*3) Scatterer-Rx (SR) link:*

$$\begin{aligned} \mathrm{g}_{SR}[n] \;&\sim\; \mathcal{CN}\Big(\sqrt{\tfrac{K_{SR}}{K_{SR}+1}}\;e^{j\,\theta_{SR}[n]},\; \tfrac{1}{K_{SR}+1}\Big) \\ &\sim\; \mathcal{CN}\big(\mu_{\mathrm{g}_{SR}}[n],\; \sigma^2_{\mathrm{g}_{SR}}\big) \end{aligned} \tag{29}$$

$$\mathbf{G}_{SR} = \mathbf{F}\,\mathbf{g}_{SR} \;\sim\; \mathcal{CN}\big(\boldsymbol{\mu}_{\mathbf{G}_{SR}},\; \boldsymbol{\Sigma}_{\mathbf{G}_{SR}}\big) \tag{30}$$

$$\boldsymbol{\mu}_{\mathbf{G}_{SR}} = [\mu_{G_{SR}}[0],\, \mu_{G_{SR}}[1],\, \ldots,\, \mu_{G_{SR}}[N-1]]^T$$

$$\mu_{\mathbf{G}_{SR}}[k] \;=\; \sum_{n=0}^{N-1} \sqrt{\frac{K_{SR}}{K_{SR}+1}}\, e^{j\,\theta_{SR}(n)}\, e^{-j\,\frac{2\pi}{N}\,k\,n}$$

$$\boldsymbol{\Sigma}_{\mathbf{G}_{SR}} = \frac{N}{K_{SR}+1}\,\mathbf{I}_N$$

*4) Cascaded path with TS and SR links:* A detailed derivation of the exact distribution, along with a workaround, is presented in Appendix A. Since the exact distribution involves the *Confluent Hypergeometric Function* and is not tractable for classifier design, we employ moment matching to approximate the distribution of the cascaded channel.

$$\mathbf{G}_{cascaded} \;\sim\; \mathcal{CN}\big(\boldsymbol{\mu}_{\mathbf{G}_{cascaded}},\; \boldsymbol{\Sigma}_{\mathbf{G}_{cascaded}}\big) \tag{31}$$

$$
\begin{aligned}
\boldsymbol{\mu}_{\mathbf{G}_{cascaded}} &= \boldsymbol{\mu}_{\mathbf{G}_{TS}} \odot \boldsymbol{\mu}_{\mathbf{G}_{SR}} \\
\boldsymbol{\Sigma}_{\mathbf{G}_{cascaded}} &= \boldsymbol{\Sigma}_{\mathbf{G}_{TS}} \boldsymbol{\Sigma}_{\mathbf{G}_{SR}} + |\mu_{G_{SR}}|^2 \boldsymbol{\Sigma}_{\mathbf{G}_{TS}} + |\mu_{G_{TS}}|^2 \boldsymbol{\Sigma}_{\mathbf{G}_{SR}} \\
&= \boldsymbol{\Sigma}_{\mathbf{G}_{TS}} \boldsymbol{\Sigma}_{\mathbf{G}_{SR}} + \frac{N^2 K_{SR}}{K_{SR}+1} \boldsymbol{\Sigma}_{\mathbf{G}_{TS}} + \frac{N^2 K_{TS}}{K_{TS}+1} \boldsymbol{\Sigma}_{\mathbf{G}_{SR}} \\
&= \frac{N^2 + N^3 K_{SR} + N^3 K_{TS}}{(K_{TS}+1)(K_{SR}+1)} \mathbf{I}_N
\end{aligned}
$$

### *E. Channels under each Hypothesis*

*1) Under $\mathcal{H}_0$:* There is no scatterer.

$$
\begin{aligned}
G_{\mathcal{H}_0} &= G_{\text{TR}} \\
&\sim \mathcal{CN}(\boldsymbol{\mu}_{\mathbf{G}_{TR}}, \boldsymbol{\Sigma}_{\mathbf{G}_{TR}}) \\
&\sim \mathcal{CN}(\boldsymbol{\mu}_{\mathcal{H}_0}, \sigma^2_{\mathcal{H}_0} \mathbf{I}_N) \\
&\sim \mathcal{CN}(\boldsymbol{\mu}_{\mathcal{H}_0}, \Sigma_{\mathcal{H}_0})
\end{aligned} \tag{32}
$$

*2) Under $\mathcal{H}_1$:* Scatterer is present, hence we will have a direct path and a scattered path.

$$
\begin{aligned}
G_{\mathcal{H}_1} &= G_{\text{TR}} + G_{\text{cascaded}} \\
&\sim \mathcal{CN}(\boldsymbol{\mu}_{\mathbf{G}_{TR}} + \boldsymbol{\mu}_{\mathbf{G}_{cascaded}}, \boldsymbol{\Sigma}_{\mathbf{G}_{TR}} + \boldsymbol{\Sigma}_{\mathbf{G}_{cascaded}}) \\
&\sim \mathcal{CN}(\boldsymbol{\mu}_{\mathcal{H}_1}, \sigma^2_{\mathcal{H}_1} \mathbf{I}_N) \\
&\sim \mathcal{CN}(\boldsymbol{\mu}_{\mathcal{H}_1}, \Sigma_{\mathcal{H}_1})
\end{aligned} \tag{33}
$$

### *F. Derivation of the Optimum Detection Rule*

We consider the binary hypothesis testing problem where the observed (complex) vector $\mathbf{G}$ is drawn from one of two distributions as given in (32) and (33).
The probability density function (pdf) for a complex Gaussian vector is

$$
p(\mathbf{G}) = \frac{1}{\pi^N \det(\boldsymbol{\Sigma})} \exp\Big[-(\mathbf{G}-\boldsymbol{\mu})^{\dagger} \boldsymbol{\Sigma}^{-1} (\mathbf{G}-\boldsymbol{\mu})\Big] \tag{34}
$$

The likelihood ratio is given by

$$
\Lambda(\mathbf{G}) = \frac{f(\mathbf{G};\mathcal{H}_1)}{f(\mathbf{G};\mathcal{H}_0)} \underset{\mathcal{H}_0}{\overset{\mathcal{H}_1}{\gtrless}} \eta \tag{35}
$$

Substituting the densities using (32), (33), and (34):

$$
\Lambda(\mathbf{G}) = \frac{\dfrac{1}{\pi^N \det(\boldsymbol{\Sigma}_{\mathcal{H}_1})} \exp\Big[-(\mathbf{G}-\boldsymbol{\mu}_{\mathcal{H}_1})^H \boldsymbol{\Sigma}^{-1}_{\mathcal{H}_1} (\mathbf{G}-\boldsymbol{\mu}_{\mathcal{H}_1})\Big]}{\dfrac{1}{\pi^N \det(\boldsymbol{\Sigma}_{\mathcal{H}_0})} \exp\Big[-(\mathbf{G}-\boldsymbol{\mu}_{\mathcal{H}_0})^H \Sigma^{-1}_{\mathcal{H}_0} (\mathbf{G}-\boldsymbol{\mu}_{\mathcal{H}_0})\Big]} \tag{36}
$$

After simplifying and taking the natural logarithm of the likelihood ratio in (36), we obtain the log-likelihood ratio (LLR):

$$
\begin{aligned}
\ln \Lambda(\mathbf{G}) = \ln\!\left(\frac{\det(\boldsymbol{\Sigma}_{\mathcal{H}_0})}{\det(\boldsymbol{\Sigma}_{\mathcal{H}_1})}\right) &- (\mathbf{G}-\boldsymbol{\mu}_{\mathcal{H}_1})^H \boldsymbol{\Sigma}^{-1}_{\mathcal{H}_1} (\mathbf{G}-\boldsymbol{\mu}_{\mathcal{H}_1}) \\
&+ (\mathbf{G}-\boldsymbol{\mu}_{\mathcal{H}_0})^H \boldsymbol{\Sigma}^{-1}_{\mathcal{H}_0} (\mathbf{G}-\boldsymbol{\mu}_{\mathcal{H}_0})
\end{aligned} \tag{37}
$$

Expanding the Quadratic Forms:

$$
\begin{aligned}
(\mathbf{G}-\boldsymbol{\mu}_{\mathcal{H}_0})^H \Sigma^{-1}_{\mathcal{H}_0} (\mathbf{G}-\boldsymbol{\mu}_{\mathcal{H}_0}) &= \mathbf{G}^H \Sigma^{-1}_{\mathcal{H}_0} \mathbf{G} \\
&\quad - 2\,\Re\{\boldsymbol{\mu}^H_{\mathcal{H}_0} \Sigma^{-1}_{\mathcal{H}_0} \mathbf{G}\} + \boldsymbol{\mu}^H_{\mathcal{H}_0} \Sigma^{-1}_{\mathcal{H}_0} \boldsymbol{\mu}_{\mathcal{H}_0} \\
(\mathbf{G}-\boldsymbol{\mu}_{\mathcal{H}_1})^H \Sigma^{-1}_{\mathcal{H}_1} (\mathbf{G}-\boldsymbol{\mu}_{\mathcal{H}_1}) &= \mathbf{G}^H \Sigma^{-1}_{\mathcal{H}_1} \mathbf{G} \\
&\quad - 2\,\Re\{\boldsymbol{\mu}^H_{\mathcal{H}_1} \Sigma^{-1}_{\mathcal{H}_1} \mathbf{G}\} + \boldsymbol{\mu}^H_{\mathcal{H}_1} \Sigma^{-1}_{\mathcal{H}_1} \boldsymbol{\mu}_{\mathcal{H}_1}
\end{aligned}
$$

Then the LLR becomes

$$
\begin{aligned}
\ln \Lambda(\mathbf{G}) = \ln\!\left[\frac{\det(\Sigma_{\mathcal{H}_0})}{\det(\Sigma_{\mathcal{H}_1})}\right] &+ \mathbf{G}^H \big(\Sigma^{-1}_{\mathcal{H}_0} - \Sigma^{-1}_{\mathcal{H}_1}\big) \mathbf{G} \\
&- 2\Re\big\{\mu^H_{\mathcal{H}_0} \Sigma^{-1}_{\mathcal{H}_0} \mathbf{G} - \mu^H_{\mathcal{H}_1} \Sigma^{-1}_{\mathcal{H}_1} \mathbf{G}\big\} \\
&+ \big(\mu^H_{\mathcal{H}_0} \Sigma^{-1}_{\mathcal{H}_0} \mu_{\mathcal{H}_0} - \mu^H_{\mathcal{H}_1} \Sigma^{-1}_{\mathcal{H}_1} \mu_{\mathcal{H}_1}\big)
\end{aligned} \tag{38}
$$

The decision rule is:

$$
\Lambda(\mathbf{G}) \underset{\mathcal{H}_0}{\overset{\mathcal{H}_1}{\gtrless}} \eta
$$

or equivalently, using the log form,

$$
\ln \Lambda(\mathbf{G}) \underset{\mathcal{H}_0}{\overset{\mathcal{H}_1}{\gtrless}} \gamma
$$

where $\eta$ (or $\gamma$) is a threshold determined by the prior probabilities and costs.

### *G. Computational Complexity*

*1) For the Likelihood Ratio Test:* We break the computation of (38) into a *preprocessing* phase and a *per-test-sample* phase.

*a) Preprocessing (one-time) costs:*

- **Matrix factorizations.** Compute Cholesky (or LU) decompositions of each $N \times N$ covariance, $\Sigma_{\mathcal{H}_i} = L_i L_i^H$. Cost:
$$2 \times \mathcal{O}\big(\tfrac{1}{3}N^3\big) \equiv \mathcal{O}(N^3).$$
- **Determinants.** Extract $\ln\det(\Sigma_{\mathcal{H}_i})$ from the diagonal of $L_i$. Cost is negligible once the factorization is done.
- **Triangular solves for means.** Compute $\mathbf{v}_i = L_i^{-1} \mu_{\mathcal{H}_i}$ to form $\Sigma^{-1}_{\mathcal{H}_i} \mu_{\mathcal{H}_i}$. Cost:
$$2 \times \mathcal{O}(N^2) \equiv \mathcal{O}(N^2).$$

*b) Per-test-sample (runtime) costs:* Given a test vector $\mathbf{G}$:

- **Triangular solves for data.** Compute $\mathbf{w}_i = L_i^{-1}\mathbf{G}$, which yields $\Sigma^{-1}_{\mathcal{H}_i}\mathbf{G}$. Cost per hypothesis: $\mathcal{O}(N^2)$, total $2\,\mathcal{O}(N^2) \equiv \mathcal{O}(N^2)$.
- **Quadratic form.** Evaluate $\mathbf{G}^H(\Sigma^{-1}_{\mathcal{H}_0} - \Sigma^{-1}_{\mathcal{H}_1})\mathbf{G}$ via two inner products of length $N$: $\mathcal{O}(N)$.
- **Cross-terms.** Compute $\mu^H_{\mathcal{H}_i} \Sigma^{-1}_{\mathcal{H}_i} \mathbf{G} = \mathbf{v}_i^H \mathbf{w}_i$, each $\mathcal{O}(N)$.
- **Constant term.** The precomputed $\mu^H_{\mathcal{H}_i} \Sigma^{-1}_{\mathcal{H}_i} \mu_{\mathcal{H}_i}$ incurs no runtime cost.

*c) Overall complexity:*

- *Preprocessing:* $\mathcal{O}(N^3)$ dominated by two Cholesky/LU factorizations.
- *Per-test-sample:* $\mathcal{O}(N^2)$ dominated by two triangular solves; all inner products are only $\mathcal{O}(N)$.

Hence, for every test vector, the overall complexity is dominated by the $\mathcal{O}(N^3)$ preprocessing, plus $\mathcal{O}(N^2)$ per evaluation, motivating PCA-based reduction to a $P \ll N$ subspace.

*2) For PCA based LRT:* After computing the principal subspace once, PCA+LRT replaces the full $N$-dimensional LRT by an $P$-dimensional test. We divide the cost into:

*a) One-time (training) costs:*

- **Eigen-decomposition (or SVD):** computing the top $P$ eigen-vectors/ singular-vectors of the $N \times N$ covariance matrix can be done in $\mathcal{O}(N^3)$ (or $\mathcal{O}(N^2P)$ with Lanczos methods). The result is the projection matrix: $U_P \in \mathbb{C}^{N\times P}$.
- **Subspace covariance inversion:** forming and decomposing the $P\times P$ projected covariance $\Sigma_P = U_P^H\,\Sigma\,U_P$ costs $\mathcal{O}(NP^2)$ to compute and $\mathcal{O}(P^3)$ to factor.

*b) Per-test-sample (inference) costs:* For each new CSI vector $\mathbf{G} \in \mathbb{C}^N$:

- **Projection:** $z = U_P^H\mathbf{G}$ requires $\mathcal{O}(NP)$ operations.
- **Triangular solves:** solving $\mathbf{\Sigma}_P^{-1}z$ via the $P\times P$ Cholesky factors costs $\mathcal{O}(P^2)$.
- **Quadratic form & constants:** computing $z^H\mathbf{\Sigma}_P^{-1}z$ and the mean-offset terms costs $\mathcal{O}(P^2)$.

Overall, each detection decision costs

$$\mathcal{O}(NP)\;+\;\mathcal{O}(P^2)\;=\;\mathcal{O}\big(NP+P^2\big),$$

versus $\mathcal{O}(N^2)$ per sample (after $\mathcal{O}(N^3)$ setup) in the full LRT.

*3) Complexity Comparison:*

$$\begin{aligned}&\text{Full LRT:} && \mathcal{O}(N^3)\ \text{(setup)}\ +\mathcal{O}(N^2)\ \text{per test-sample},\\ &\text{PCA-LRT:} && \mathcal{O}(N^3)\ \text{(PCA setup)}\ +\mathcal{O}(NP+P^2)\ \text{per test-sample}.\end{aligned}$$

When $P \ll N$, the per-sample cost drops from $\mathcal{O}(N^2)$ to $\mathcal{O}(NP)$, and the small $P$-dimensional inversion $\mathcal{O}(P^3)$ is a negligible one-time cost.

### *H. Advantages (and Caveats) of Using PCA for Dimension Reduction*

- **Computational Efficiency:** By projecting the data onto a lower-dimensional subspace (say, $P \ll N$) via PCA, the test statistic computation becomes more manageable. Inverting $P\times P$ matrices and computing the determinants in that lower-dimensional space is computationally faster and numerically more stable.
- **Nullifying Inter-dimensional Correlation:** PCA diagonalizes the covariance matrix, transforming the original, correlated feature space into a new set of orthogonal (i.e. uncorrelated) principal components. By "nulling out" inter-dimension correlations, PCA not only simplifies the statistical structure so that each retained component carries independent information but also improves the stability and efficiency of downstream algorithms (e.g. Gaussian LRT or SVM), which typically assume uncorrelated inputs.
- **Noise Reduction:** PCA is often used as a noise-reduction tool. If the discriminative information is concentrated in only a few dimensions (e.g., as a rank-one or low-rank update in the covariance), then focusing on those few dimensions can improve the signal-to-noise ratio (SNR).
- **Preservation of Discriminative Information:** If the primary difference between the hypotheses is in the mean (or a structured component) and it lies in a subspace of dimension $P$, then PCA can capture most of the relevant differences. In other words, the projection $z = U_P^H x$ retains the key statistical properties that separate $H_0$ and $H_1$ (it will be discussed in Sec.III-I2).
- **Trade-off Between Complexity and Information Loss:** The advantage of reducing dimensionality is evident in lower computational complexity and improved conditioning of matrix inversions. However, this comes at the potential risk of losing some information relevant to distinguishing between the hypotheses. The ideal choice of $P$ is one that retains most of the discriminative energy (as captured, for example, by a high Bhattacharyya distance) while reducing the computational burden.

Thus, while solving the full test statistic in the original $N$-dimensional space presents challenges, PCA offers a practical way to overcome these difficulties.

### *I. Singular Value Decomposition and PCA*

We collect $N$ samples across subcarriers to form complex-valued data matrices under each hypothesis:

$$X_0 = \left[\hat{G}^{(0)}(1), \hat{G}^{(0)}(2), \ldots, \hat{G}^{(0)}(M)\right] \tag{39}$$

$$X_1 = \left[\hat{G}^{(1)}(1), \hat{G}^{(1)}(2), \ldots, \hat{G}^{(1)}(M)\right] \tag{40}$$

where $\hat{G}^{(k)}(i)$ is a $N \times 1$ channel estimate vector for each of the subcarriers on $i^{th}$ OFDM symbol under hypothesis $\mathcal{H}_k$. Each $X_k$ is an N × M matrix, where M is the number of OFDM symbols.

We form the overall data matrix $\mathbf{X} = [X_0\ X_1]\ \in \mathbb{C}^{N\times M_{\text{tot}}}$ (with $M_{\text{tot}} = 2M$ and the classes have equal priors).

*1) SVD of Data Matrices:* We then perform a Singular Value Decomposition (SVD) of $\mathbf{X}$:

$$\mathbf{X} = \mathbf{U}\,\mathbf{\Sigma}\,\mathbf{V}^H,$$

where:

- $\mathbf{U} \in \mathbb{C}^{N\times N}$ contains the left singular vectors (the principal directions).
- $\mathbf{\Sigma} \in \mathbb{R}^{N\times M_{\text{tot}}}$ is a (rectangular) diagonal matrix containing the singular values $\sigma_1 \geq \sigma_2 \geq \cdots \geq \sigma_N$.
- $\mathbf{V} \in \mathbb{C}^{M_{\text{tot}}\times M_{\text{tot}}}$ contains the right singular vectors.

By choosing the top $P$ singular values, we form the projection matrix

$$\mathbf{U}_P = [\,\mathbf{u}_1 \quad \mathbf{u}_2 \quad \cdots \quad \mathbf{u}_P\,] \in \mathbb{C}^{N\times P}.$$

Defining

$$\alpha = \frac{\sigma^2_{\mathcal{H}_0} + \sigma^2_{\mathcal{H}_1}}{2}, \quad \beta = \frac{1}{4}, \quad \mathbf{u} = \boldsymbol{\mu}_{\mathcal{H}_0} - \boldsymbol{\mu}_{\mathcal{H}_1}$$

As derived in Appendix B, from (47), the overall covariance of $X$ can be given by,

$$\boxed{\Sigma_{\mathbf{X}} = \alpha\, I_N + \beta\,\mathbf{u}\,\mathbf{u}^H} \tag{41}$$

*2) Projection onto Principal Components:* The PCA projection of any sample $x \in \mathbf{X}$ is then given by

$$z = \mathbf{U}_P^H x \quad \in \mathbb{C}^P.$$

- The projected data $z$ represents the data in the new coordinate system defined by the principal components, effectively reducing dimensionality while retaining significant variance.
- (41) shows that the covariance of the overall data remains an identity-scaled term (can be thought of as noise floor) plus a rank-one update that captures the structured difference in the means.

### *J. Upper limit on $P_e$ Using Bhattacharyya Bound*

The Bhattacharyya bound [20], [21] relates the statistical properties of the projected data clusters under each hypothesis to the classification error probability.

From (48) and (49) in Appendix C, for equal priors, the probability of error $P_e$ is bounded by

$$P_e \leq \frac{1}{2} e^{-D_B}$$

where

$$D_B = \frac{1}{4(\sigma^2_{\mathcal{H}_0} + \sigma^2_{\mathcal{H}_1})} \| U_P^H (\boldsymbol{\mu}_{\mathcal{H}_0} - \boldsymbol{\mu}_{\mathcal{H}_1}) \|^2 + \frac{P}{2} \ln \left( \frac{\sigma^2_{\mathcal{H}_0} + \sigma^2_{\mathcal{H}_1}}{2\sigma_{\mathcal{H}_0}\sigma_{\mathcal{H}_1}} \right) \tag{42}$$

Thus, the **upper bound on the classification accuracy** is given by

$$1 - P_e \geq 1 - \frac{1}{2} e^{-D_B}.$$

A higher Bhattacharyya distance $D_B$ implies a lower error probability and, therefore, higher classification accuracy.

## IV. Simulation Results

### *A. Aim:*

The primary goal of our simulations is to demonstrate the trade-off between detection performance and computational complexity when using OFDM-signal and PCA-based dimensionality reduction for scatterer sensing. By comparing the Full LRT against three reduced-rank variants (PCA+LRT, PCA+Linear SVM, PCA+RBF SVM), and sweeping both the PCA subspace dimension $P$ and the receive SNR, we seek to demonstrate:

- **Computational savings:** The order-of-magnitude reduction in inference time (post-training) achieved by working in a $P$-dimensional subspace rather than the full $N$-dimensional space.
- **Robustness to dimension:** The behavior of all detectors at two representative data dimensions, $N = 256$ and $N = 1024$, highlighting the scalability of PCA methods in very high dimensions.
- **Accuracy retention:** How rapidly each SVM-based detectors approach the LRT error rate as $P$ increases.
- **Noise resilience:** The sensitivity of each detector's performance across a wide SNR range (from $-10$dB to $+15$dB), illustrating when a small PCA subspace suffices for near Bayes-optimal detection.

Together, these experiments validate that a low-rank PCA subspace ($P \ll N$) can preserve almost all of the scatterer signature enabling real-time, high-accuracy sensing in next-generation wireless systems.

### *B. Computational Complexity vs. PCA Dimension*

Fig. 4 reports the *post-training* inference time (log-scale) at SNR = 5dB as we sweep the PCA subspace dimension $P$, for $N = 256$ (left) and $N = 1024$ (right).

- **Full LRT** (solid red) independent of $P$, so it remains flat in $P$. Since it inverts the full $N \times N$ covariance matrix. This costs roughly 2s for $N = 256$ and about 17s for $N = 1024$.
- **PCA+LRT** (blue circles) inverts only a $P \times P$ covariance. As $P$ grows to 80, its decision time stays between $0.09 - 0.1$s and $1 - 2$s, for $N = 256$ and $N = 1024$ respectively.
- **For the SVM based methods**, Because at very low P the SVM's fixed overhead (e.g. data structure setup, kernel-matrix allocation, interpreter overhead) dominates the tiny cost of classifying a handful of features. Adding a few more dimensions lets those one time costs be amortized over more computations, so the measured runtime actually *drops* as P increases from 1 to roughly 10.
- **PCA+SVM (Linear and RBF Kernels)** projects the test data into $P$ dimensions then applies a SVM with specified kernel. Its inference cost is extremely low, almost similar to the PCA+LRT method: Its decision time stays between $0.09 - 0.1$s and $1 - 2$s, for $N = 256$ and $N = 1024$ respectively.
- **Practical Implication:** By combining a small PCA subspace ($P \lesssim 20$) with a lightweight SVM classifier, CommSense achieves small inference time, order of magnitude faster than the Full LRT.

### *C. Classification Error vs. PCA Dimension*

Fig. 5 shows the classification error as a function of PCA subspace dimension $P$ for two data dimensions ($N = 256$ and $N = 1024$) and three SNR levels (0, 5, 10dB). We compare four detectors: Full LRT (red line), PCA+LRT (blue circles), PCA+SVM linear (green squares), and PCA+SVM RBF (magenta diamonds) and overlay the Bhattacharyya bound after PCA (dashed black).

**Figures 5a-5c $N = 256$:**

- *Full LRT* achieves zero error across all SNRs, confirming that the full $256 \times 256$ covariance inversion perfectly separates the hypotheses.
- *PCA+LRT* stays below the Bhattacharyya bound: at SNR = 0dB, error drops below $10^{-2}$ by $P \approx 40$; at SNR $\geq$ 5dB, only $P \approx 10 - 15$ is needed.
- *PCA+SVM* (linear and RBF) closely follow the PCA based Bhattacharya Bound at SNR=0dB. Both the kernels briefly outperforms the bound at $P \approx 10$ (e.g. SNR=5dB), but adding more components beyond the

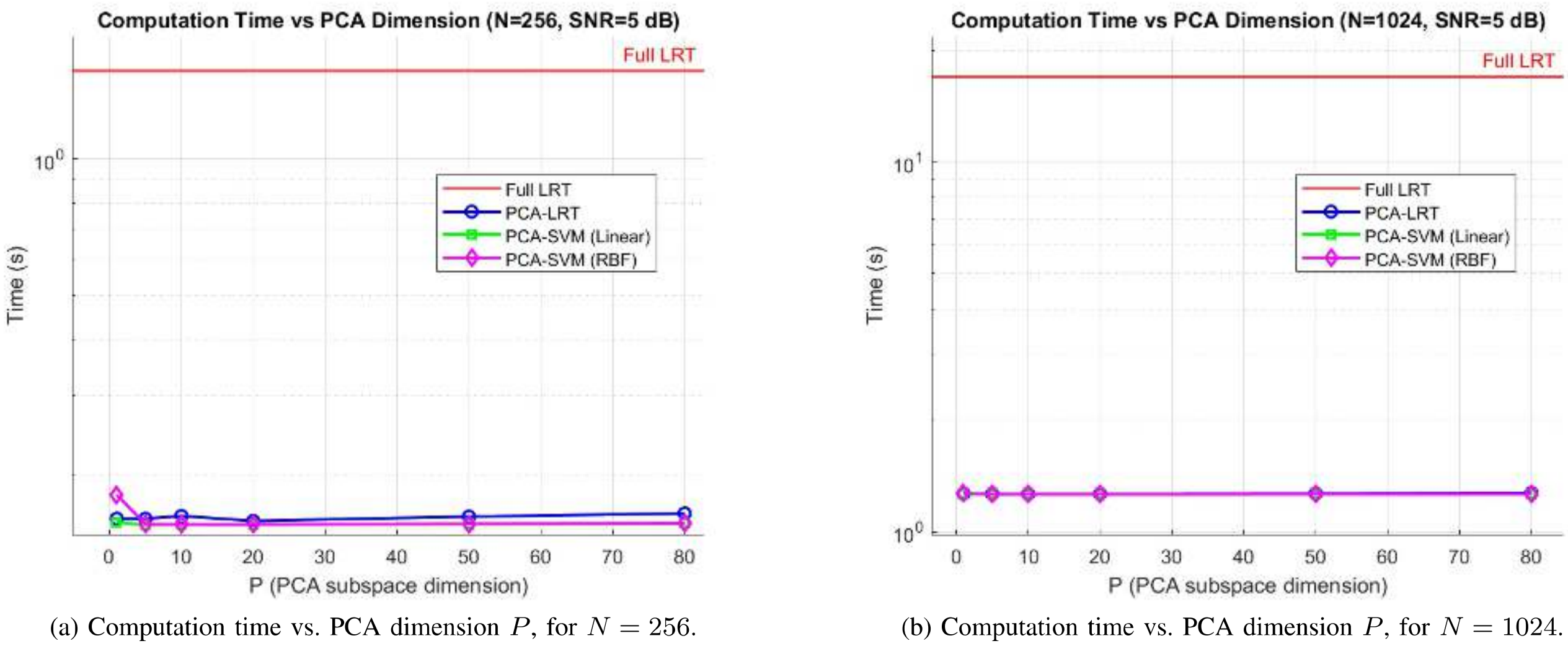

(a) Computation time vs. PCA dimension $P$, for $N = 256$.

(b) Computation time vs. PCA dimension $P$, for $N = 1024$.

Fig. 4. Post-training inference time as a function of PCA subspace size. Each subplot reports the compute time (in seconds, log scaled) for four detectors: the full dimensional LRT (solid red line), PCA+LRT (blue circles), PCA+SVM (linear kernel, green squares), and PCA+SVM (RBF kernel, magenta diamonds), as the number of retained principal components $P$ increases from 1 to 80.

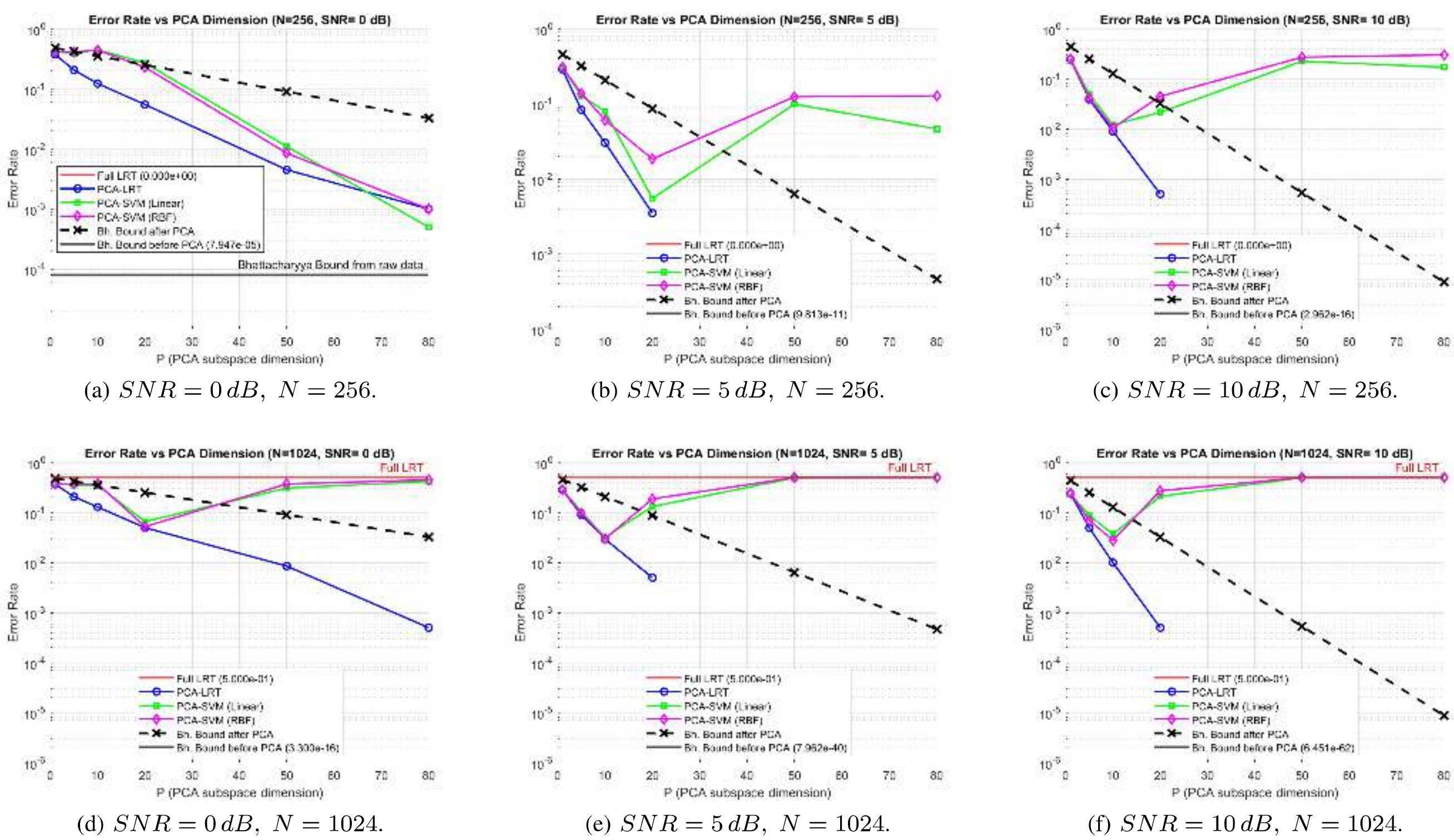

(a) $SNR = 0\,dB,\ N = 256$.

(b) $SNR = 5\,dB,\ N = 256$.

(c) $SNR = 10\,dB,\ N = 256$.

(d) $SNR = 0\,dB,\ N = 1024$.

(e) $SNR = 5\,dB,\ N = 1024$.

(f) $SNR = 10\,dB,\ N = 1024$.

Fig. 5. Error rate vs. PCA dimension $P$. We compare four detectors: Full LRT (red), PCA–LRT (blue circles), PCA+SVM linear (green squares), and PCA+SVM RBF (magenta diamonds) against the theoretical Bhattacharyya bound computed from the raw data (solid gray) and after PCA (dashed black).

"elbow" injects noise-dominated features and causes a error uptick.

**Figures 5d-5f $N = 1024$:**

- *Full LRT* fails completely (error $\approx 0.5$) at all SNRs, since inverting a $1024 \times 1024$ covariance causes numerical overflow, it forces the error to collapse toward 0.5 (random guessing), so it no longer appears as a meaningful detector.
- *PCA+LRT* again stays well below the bound: at SNR=0dB, one needs $P \approx 45$ to drive error below $10^{-2}$; at SNR=5dB and 10dB, $P \gtrsim 10$ suffices.
- *PCA+SVM* again tracks the bound; Linear SVM achieves error $\approx$0.03 by $P \approx 10$ at SNR $\geq$5dB. The RBF SVM dips slightly below linear at moderate $P$ but also degrades for $P > 20$ due to noisy components.

Overall, these results confirm that projecting high-dimensional CSI onto a low-rank PCA subspace ($P \ll N$) and then applying either a reduced-rank LRT or a lightweight SVM yields performance that closely approaches the Bayes-optimal bound, while the full-dimensional LRT becomes computationally prohibitive at large $N$.

### D. Impact of Mean/Covariance Estimation Error on LRT

To evaluate the robustness of our reduced rank detectors to imperfect parameter knowledge, we injected some perturbations into the estimated class means and covariance matrices prior to applying the LRT decision rule. Figs. 6a and 6b show the resulting test-set error rates as a function of the percentage error in both $\boldsymbol{\mu}$ and $\Sigma$, for $N = 256$ subcarriers at SNR=0dB and 5dB, respectively. We plot curves for PCA+LRT with $P \in \{1, 5, 10, 20\}$ components and include the Full LRT benchmark (black $\times$).

- **Full LRT robustness:** With perfect parameter estimates, Full LRT is Bayes-optimal and error-free. Even under severe perturbations ($\leq 30\%$), it maintains zero error, highlighting its theoretical resilience but practical vulnerability at large $N$.
- **PCA-LRT sensitivity to $P$:** For $P = 1$, the error remains high ($\approx$0.37–0.30) regardless of SNR or estimation error. Moderate subspaces ($P = 5, 10, 20$) offer a trade-off: they reduce baseline error but causes gradual performance degradation as estimation error grows.
- **Practical implication:** In practice, estimation errors from finite training data undermine PCA+LRT's practicality, despite large subspaces recovering near-optimal performance. So we favor PCA+SVM, which requires no parameter estimates.

### E. ROC Curves vs. SNR

Finally, Fig. 7 shows ROC curves for detecting scatterers at six SNR settings for across two data dimensions ($N = 256$ and $N = 1024$). Here we are keeping the PCA subspace dimension (P) fixed at 10.

Table II summarizes the Area Under ROC curve (AUC) for each of the detectors.

**Low-SNR regime ($-10$dB to 0dB):** For $N = 256$, the Full LRT already achieves a high AUC (0.92 at $-10$dB, 1.00 at $-5$dB), whereas PCA+LRT and PCA+SVM variants lagged (AUC 0.59-0.61 at $-10$dB). By 0dB, however, PCA+LRT climbs to 0.95 and PCA+SVM to 0.93-0.94, nearly matching Full LRT. In contrast, for $N = 1024$ the Full LRT performs poorly. Remarkably, PCA+LRT and both PCA+SVM methods already exceed 0.60 AUC at $-10$dB, rising above 0.93 by 0dB.

**High-SNR regime (5dB to 15dB):** At 5dB and above, for $N = 256$ all methods for both $N$ converge to near perfect discrimination (AUC $\approx$ 1.00). PCA+LRT stabilizes around 0.99, and PCA+SVM linear even reaches unity AUC for both $N$. Thus once the noise level is moderate, a small PCA subspace ($P = 10$) suffices to achieve virtually optimal detection performance, regardless of the original signal dimension.

### F. Discussion

Our simulation study demonstrates that PCA-based dimensionality reduction unlocks a powerful trade off between computational efficiency and detection performance in high dimensional OFDM signal based scatterer sensing.

First, as shown in Figs. 4a-4b, projecting the $N$-dimensional channel measurements onto a low-rank subspace of size $P \approx 10$ yields a order of magnitude reduction in inference time relative to the full dimensional LRT. In practical terms, making real-time deployment feasible even for large subcarrier counts.

Crucially, this speedup comes with almost no loss in detection accuracy. Table II show that PCA+LRT matches the full LRT's ROC AUC once $P \sim 10$, and both PCA+SVM (linear) and PCA+SVM (RBF) converge to optimal performance with only modest extra cost. In other words, essentially all of the scatterer "signature" is captured in the first few principal components, and one can safely ignore the remaining dimensions without hurting accuracy.

This effect is even more pronounced in very high dimensions. For $N = 1024$, the full LRT breaks down due to numerical instability, whereas our PCA based detectors already achieve $AUC \approx 0.6$ at $-10\,$dB and exceed 0.90 by $0\,$dB (Table II). By concentrating the "energy" of the scatterer signal into the top singular-vectors, PCA acts as an effective noise filter, enabling robust detection even when most dimensions are meaningless.

In practical settings, CSI estimates and covariance matrices must be learned from finite length training data, so parameter mismatch is unavoidable. While PCA+LRT can recover near optimal performance by retaining enough principal components, its reliance on accurate mean and covariance estimates makes it increasingly fragile as estimation error grows. In contrast, PCA+SVM classifiers operate directly on the reduced dimension features without requiring explicit parameter inversion, and therefore exhibit far greater resilience to model uncertainty: making them the preferred choice when facing non ideal scenarios.

Furthermore, the close alignment between our empirical error curves and the PCA based Bhattacharyya bound confirms that the leading principal directions capture virtually all the discriminative signal energy. This theoretical tightness, together with the speedups observed for PCA+SVM (see Section IV-B), supports a practical two stage architecture: first compress the high dimensional CSI into a small PCA subspace, then apply a lightweight SVM. The result is a real time detector that achieves near-Bayes-optimal accuracy at a tiny fraction of the computational cost of a full LRT.

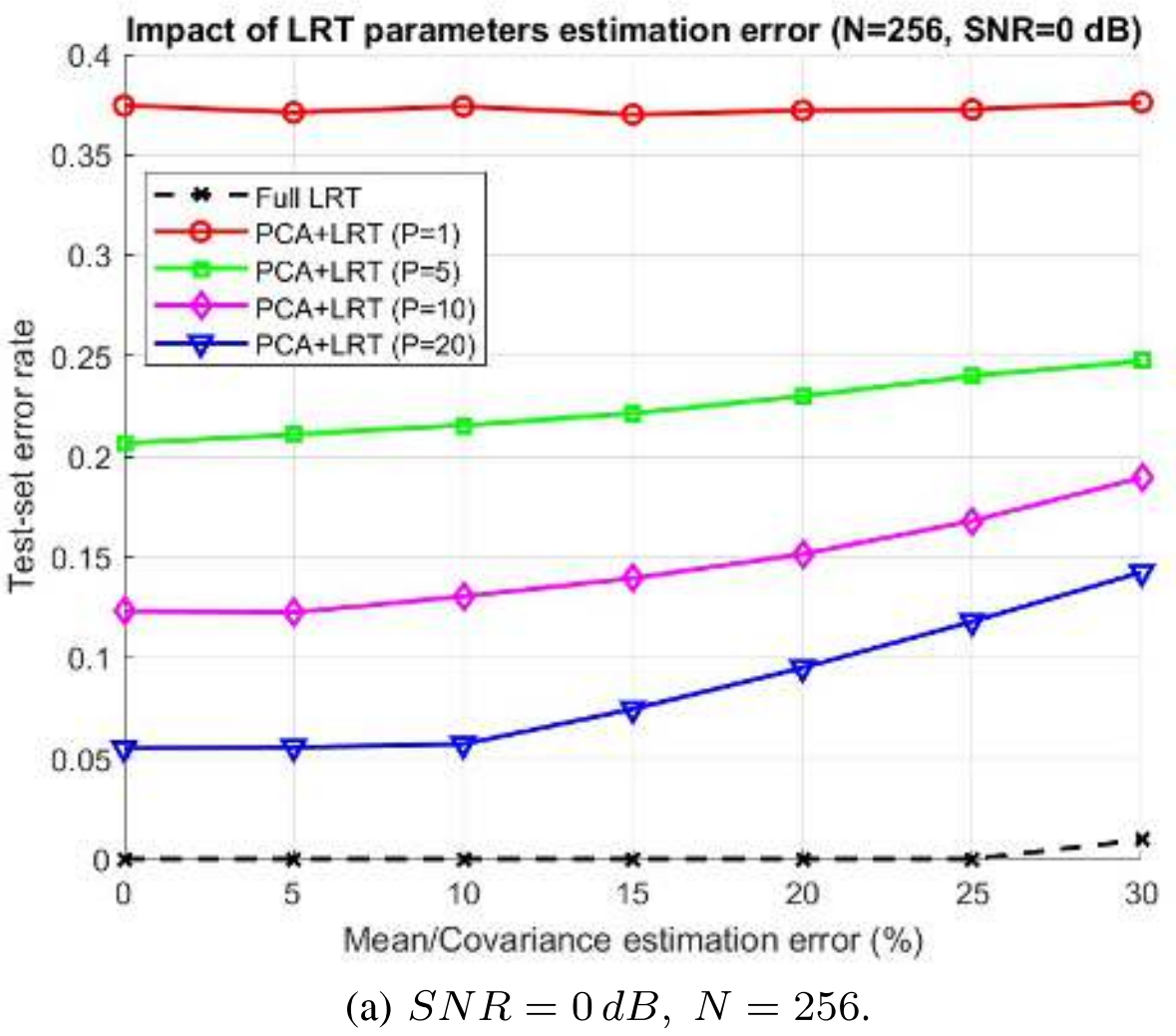

(a) $SNR = 0\,dB,\ N = 256.$

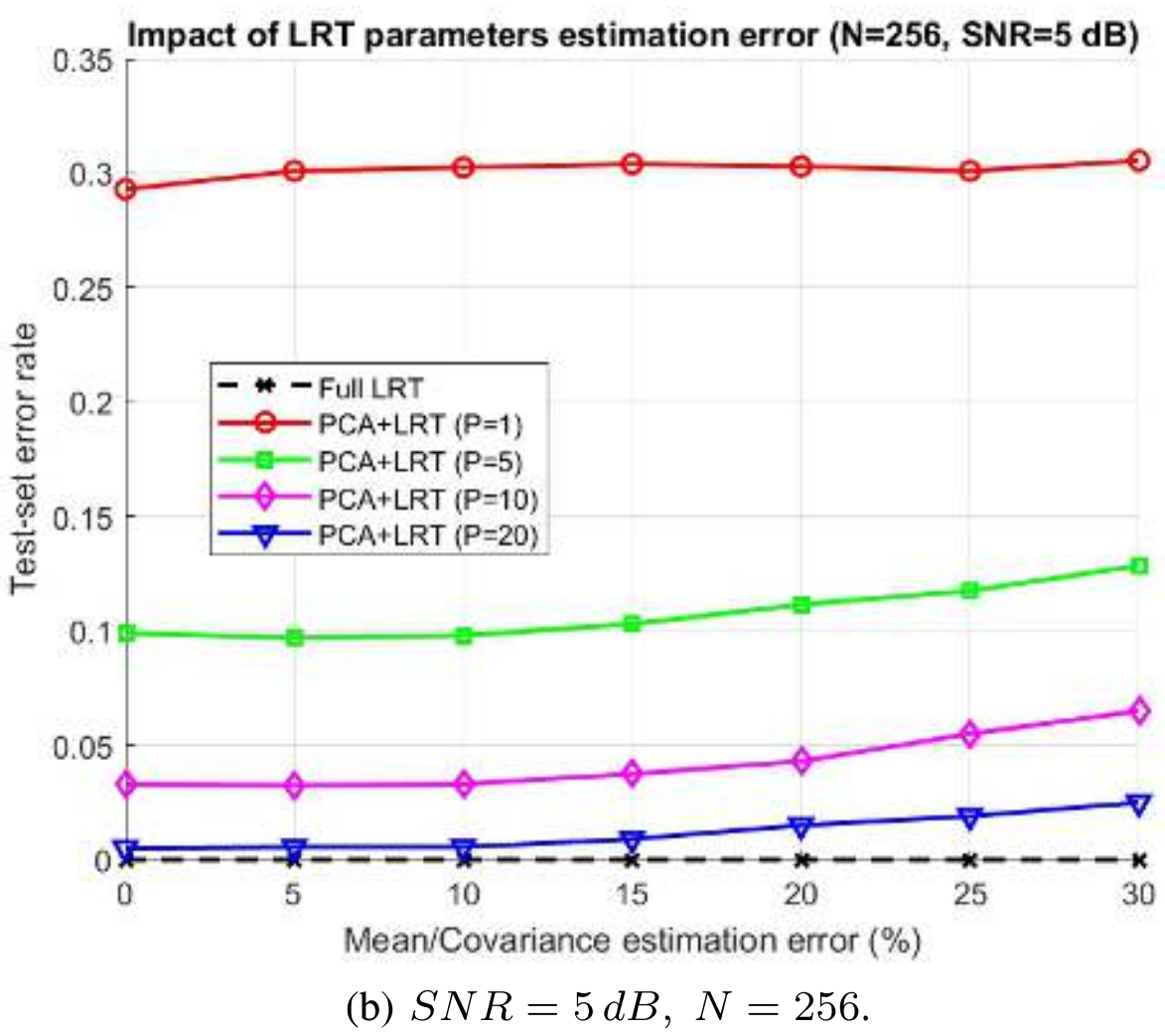

(b) $SNR = 5\,dB,\ N = 256.$

Fig. 6. **Error vs. LRT parameters perturbation ($N = 256$).** At SNR=0dB, Full LRT remains error-free up to $\sim 25\%$ mismatch. PCA–LRT with $P = 1$ is stuck at $\approx 0.37$ error, while $P = 5$ and $P = 10$ start at $\{0.20, 0.12\}$ and rise modestly to $\{0.25, 0.19\}$ at 30% error. The $P = 20$ curve is most robust, increasing only from 0.05 to 0.14. At SNR=5dB, Full LRT again tolerates $\sim 30\%$ perturbation with zero error. PCA–LRT($P = 1$) holds near 0.30, $P = 5$ climbs from 0.10 to 0.13, $P = 10$ from 0.03 to 0.07, and $P = 20$ stays below 0.02 even at 30% mismatch.

TABLE II
ROC AUC VS. SNR FOR $N = 256$ AND $N = 1024$ SUBCARRIERS ($P = 20$).

| SNR (dB) | $N = 256$ | | | | $N = 1024$ | | | |
|---|---|---|---|---|---|---|---|---|
| | Full LRT | PCA+LRT | PCA+SVM Lin | PCA+SVM RBF | Full LRT | PCA+LRT | PCA+SVM Lin | PCA+SVM RBF |
| $-10$ | 0.920 | 0.592 | 0.611 | 0.585 | 0.492 | 0.604 | 0.617 | 0.601 |
| 0 | 1.000 | 0.951 | 0.936 | 0.927 | 0.495 | 0.952 | 0.928 | 0.936 |
| 5 | 1.000 | 0.991 | 0.992 | 0.992 | 0.502 | 0.992 | 0.996 | 0.994 |
| 15 | 1.000 | 0.951 | 1.000 | 0.999 | 0.500 | 0.953 | 1.000 | 1.000 |

## V. CONCLUSION

In this paper,

- We established a rigorous theoretical analysis for the CommSense measurement instrument, enabling accelerated scatterer measurement using high-dimensional OFDM CSI with near-Bayes-optimal accuracy.
- We employed a three-link Rician measurement model (direct Tx→Rx, Tx→Scatterer, Scatterer→Rx) to simulate realistic scenarios.
- We compared four detector architectures as measurement algorithms:
  - **Full-dimensional LRT**: Statistically optimal reference method, but incurs $\mathcal{O}(N^2)$ inference complexity and possible numerical instability in higher dimensions.
  - **PCA+LRT**: Reduces instrumentation load by projecting CSI onto a $P$-dimensional subspace before applying LRT.
  - **PCA+Linear SVM**: Implements a lightweight, linear classification module on PCA features for real-time scatterer sensing.
  - **PCA+RBF SVM**: Leverages nonlinear separation in the PCA domain to enhance measurement robustness.

Our simulation results for $N = 256$ and $N = 1024$ subcarriers over SNRs from $-10\,\text{dB}$ to $15\,\text{dB}$ revealed that:

- *Computational Efficiency:* PCA reduces inference time by an order of magnitude relative to the Full LRT (Figs. 4a–4b), making real-time measurement feasible even for very large $N$.
- *Near-Optimal Accuracy:* At $\text{SNR} \geq 0\,\text{dB}$, retaining only $P \approx 10$–$20$ principal components suffices for PCA+LRT and PCA+SVM to nearly match the Full-LRT performance based on the ROC AUC (Figs. 7a–7b, Table II).
- *High-Dimensional Robustness:* In the $N = 1024$ scenario where Full LRT fails due to numerical overflow (AUC$\approx 0.5$), PCA-based detectors achieve AUC$\gtrsim 0.60$ at $-10\,\text{dB}$ and exceed $0.90$ by $0\,\text{dB}$, illustrating that PCA inherently filters out noise-dominated directions.
- *Parameter Mismatch Tolerance:* Under up to 30% perturbation of mean/covariance estimates, PCA+LRT with larger $P$ degrades gracefully (Section IV-D), while PCA+SVM remains virtually unaffected by estimation inaccuracies thanks to its model-agnostic classifier.

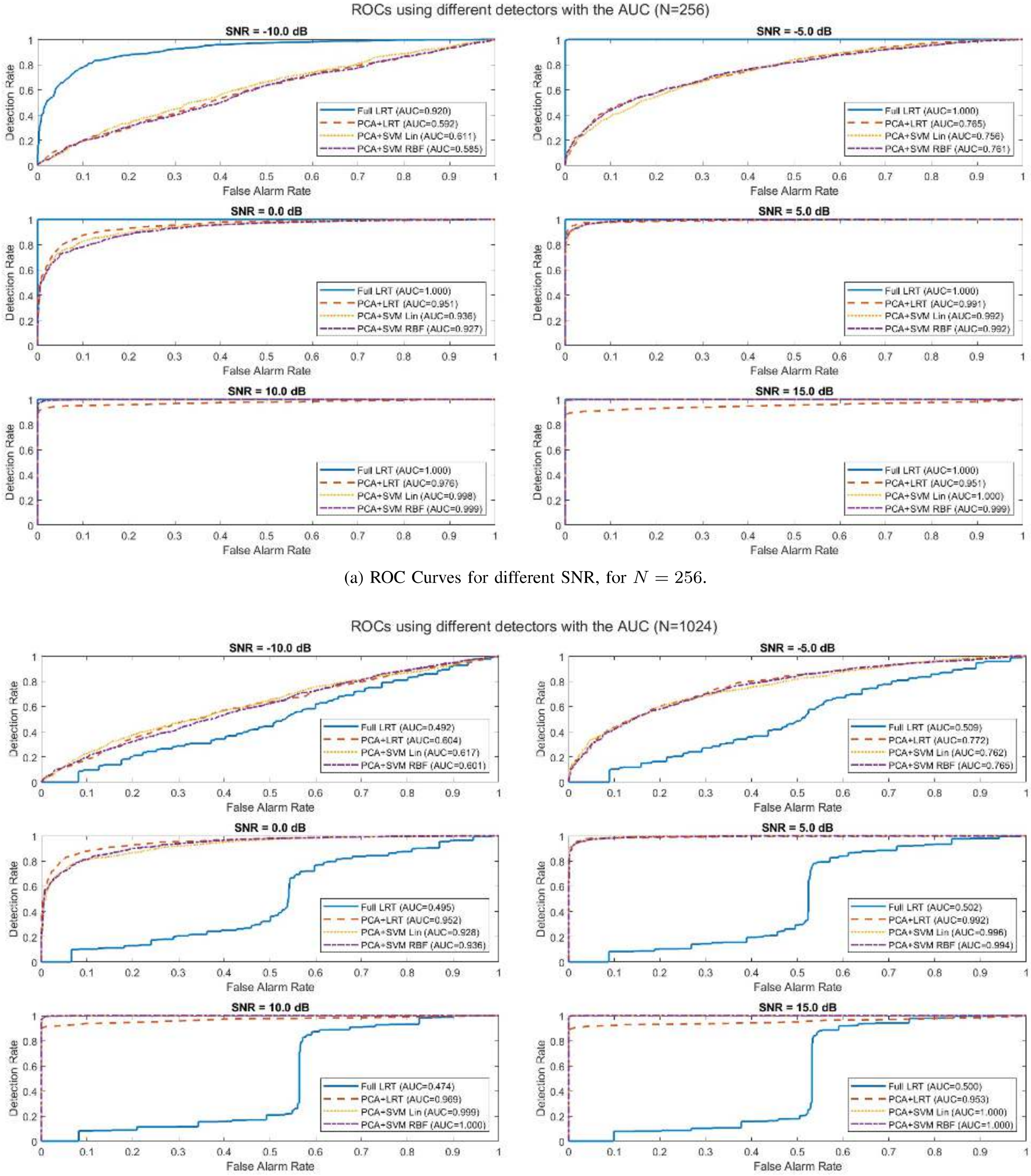

(a) ROC Curves for different SNR, for $N = 256$.

(b) ROC Curves for different SNR, for $N = 1024$.

Fig. 7. We compare four detectors: Full LRT (solid blue), PCA+LRT (dashed orange), PCA+SVM (linear, dotted yellow), and PCA+SVM (RBF, dash-dotted purple); across six SNR levels $\{-10,\ -5,\ 0,\ 5,\ 10,\ 15\,\text{dB}\}$. Each subplot overlays the ROC for all four methods at a given SNR, with the corresponding area-under-curve (AUC) shown in the legend.

- *Theoretical Consistency:* Empirical errors track the Bhattacharyya bound after PCA, confirming that the retained principal subspace indeed captures the dominant discriminative energy.

*a) Final Remarks:* The combination of PCA for dimensionality reduction and SVM for discriminative classification yields a measurement architecture that is simultaneously *fast*, *accurate*, and *robust*. By closing the gap between statistical optimality and real-time feasibility, our PCA-driven CommSense framework paves the way for practical, large-scale OFDM-based sensing in 5G/6G and beyond.

*b) Future Work:* Promising directions include:

- **Hardware Prototyping:** Build an IoT-based testbed to validate simulation findings in real propagation environments.
- **Multi-Class Sensing:** Extend to identify multiple scatterer types (e.g. vehicles, drones, humans) via one-vs-rest PCA+SVM classifiers.
- **Adaptive Subspace Tracking:** Implement online PCA updates to adapt to nonstationary channels and dynamic clutter.
- **Joint Waveform Design:** Investigate tailored OFDM pilot patterns within an ISAC framework to maximize sensing-relevant subspace energy.

## Appendix A
## Distribution of pointwise product of two non-central Complex Gaussian PDFs arising from Cascaded Channel

### A. The Complex Gaussian PDFs

Assume

$$x_1 \sim \mathcal{CN}(\mu_1, \sigma_1^2), \quad x_2 \sim \mathcal{CN}(\mu_2, \sigma_2^2),$$

with

$$f_{x_1}(x) = \frac{1}{\pi\sigma_1^2} \exp\Bigl(-\frac{|x-\mu_1|^2}{\sigma_1^2}\Bigr)$$
$$f_{x_2}(x) = \frac{1}{\pi\sigma_2^2} \exp\Bigl(-\frac{|x-\mu_2|^2}{\sigma_2^2}\Bigr)$$

Our goal is to derive the PDF of

$$z = x_1 x_2$$

### B. Transformation of Variables

Define the transformation

$$u = x_1, \quad v = x_1 x_2 \quad (v = z).$$

Then the inverse is

$$x_1 = u, \quad x_2 = \frac{z}{u},$$

and the Jacobian (in the complex plane) is

$$J = \frac{1}{|u|^2}.$$

Thus, the joint density of $(u, z)$ is

$$f_{u,z}(u, z) = f_{x_1}(u)\, f_{x_2}\Bigl(\frac{z}{u}\Bigr) \frac{1}{|u|^2}.$$

Marginalizing over $u$, we have

$$\boxed{f_z(z) = \frac{1}{\pi^2\sigma_1^2\sigma_2^2} \int_{\mathbb{C}} \exp\Biggl[-\frac{|u-\mu_1|^2}{\sigma_1^2} - \frac{\left|\frac{z}{u}-\mu_2\right|^2}{\sigma_2^2}\Biggr] \frac{d^2u}{|u|^2}.}$$

### C. Change to Polar Coordinates

Write

$$u = re^{i\theta}, \quad d^2u = r\,dr\,d\theta, \quad r \geq 0,\ \theta \in [0, 2\pi).$$

Also express

$$\mu_1 = |\mu_1|e^{i\phi_1}, \quad \mu_2 = |\mu_2|e^{i\phi_2}, \quad z = |z|e^{i\varphi}.$$

Then,

$$|u-\mu_1|^2 = r^2 - 2r|\mu_1|\cos(\theta-\phi_1) + |\mu_1|^2,$$

and

$$\left|\frac{z}{u}-\mu_2\right|^2 = \frac{|z|^2}{r^2} - 2\frac{|z|}{r}|\mu_2|\cos\Bigl(\theta-(\varphi-\phi_2)\Bigr) + |\mu_2|^2.$$

Hence, the density becomes

$$f_z(z) = \frac{1}{\pi^2\sigma_1^2\sigma_2^2} \int_0^\infty \frac{dr}{r} \int_0^{2\pi} \exp\Biggl[ - \frac{r^2 - 2r|\mu_1|\cos(\theta-\phi_1) + |\mu_1|^2}{\sigma_1^2} - \frac{\frac{|z|^2}{r^2} - 2\frac{|z|}{r}|\mu_2|\cos\bigl(\theta-(\varphi-\phi_2)\bigr) + |\mu_2|^2}{\sigma_2^2}\Biggr] d\theta$$

The $\theta$–independent factors

$$\exp\Bigl(-\frac{|\mu_1|^2}{\sigma_1^2} - \frac{|\mu_2|^2}{\sigma_2^2}\Bigr)$$

can be taken outside the integrals:

$$\boxed{f_z(z) = \frac{e^{-\frac{|\mu_1|^2}{\sigma_1^2} - \frac{|\mu_2|^2}{\sigma_2^2}}}{\pi^2\sigma_1^2\sigma_2^2} \int_0^\infty \frac{dr}{r} \exp\Bigl(-\frac{r^2}{\sigma_1^2} - \frac{|z|^2}{\sigma_2^2 r^2}\Bigr) I(r)} \tag{43}$$

with

$$I(r) = \int_0^{2\pi} \exp\Biggl[\frac{2r|\mu_1|}{\sigma_1^2}\cos(\theta-\phi_1) + \frac{2|z||\mu_2|}{\sigma_2^2 r}\cos\Bigl(\theta-(\varphi-\phi_2)\Bigr)\Biggr] d\theta$$

### D. Angular Integration via Fourier Expansion

Recall the generating function of modified Bessel functions:

$$\exp\bigl(\alpha\cos\theta\bigr) = \sum_{n=-\infty}^{\infty} I_n(\alpha)e^{in\theta}.$$

Thus,

$$\exp\Bigl(\frac{2r|\mu_1|}{\sigma_1^2}\cos(\theta-\phi_1)\Bigr) = \sum_{m=-\infty}^{\infty} I_m\Bigl(\frac{2r|\mu_1|}{\sigma_1^2}\Bigr)e^{im(\theta-\phi_1)}$$

and

$$\exp\Big(\frac{2|z||\mu_2|}{\sigma_2^2 r}\cos\big(\theta-(\varphi-\phi_2)\big)\Big) = \sum_{n=-\infty}^{\infty} I_n\Big(\frac{2|z||\mu_2|}{\sigma_2^2 r}\Big). \; e^{in(\theta-(\varphi-\phi_2))}$$

Their product is

$$\begin{aligned}
&\exp\Big[\frac{2r|\mu_1|}{\sigma_1^2}\cos(\theta-\phi_1)+\frac{2|z||\mu_2|}{\sigma_2^2 r}\cos\big(\theta-(\varphi-\phi_2)\big)\Big] \\
&= \sum_{m,n=-\infty}^{\infty} I_m\Big(\frac{2r|\mu_1|}{\sigma_1^2}\Big) I_n\Big(\frac{2|z||\mu_2|}{\sigma_2^2 r}\Big). \; e^{im(\theta-\phi_1)}e^{in(\theta-(\varphi-\phi_2))} \\
&= \sum_{m,n=-\infty}^{\infty} I_m\Big(\frac{2r|\mu_1|}{\sigma_1^2}\Big) I_n\Big(\frac{2|z||\mu_2|}{\sigma_2^2 r}\Big). \; e^{i(m+n)\theta}e^{-i\big(m\phi_1+n(\varphi-\phi_2)\big)}.
\end{aligned}$$

Integrate with respect to $\theta$:

$$\begin{aligned}
I(r) &= \int_0^{2\pi}\sum_{m,n}(\cdots)\,d\theta \\
&= \sum_{m,n} I_m\Big(\frac{2r|\mu_1|}{\sigma_1^2}\Big) I_n\Big(\frac{2|z||\mu_2|}{\sigma_2^2 r}\Big). \; e^{-i\big(m\phi_1+n(\varphi-\phi_2)\big)}\int_0^{2\pi} e^{i(m+n)\theta}d\theta
\end{aligned}$$

Since

$$\int_0^{2\pi} e^{i(m+n)\theta}d\theta = 2\pi\delta_{m,-n}$$

only terms with $n=-m$ contribute:

$$I(r) = 2\pi\sum_{m=-\infty}^{\infty} I_m\Big(\frac{2r|\mu_1|}{\sigma_1^2}\Big) I_m\Big(\frac{2|z||\mu_2|}{\sigma_2^2 r}\Big)e^{im\big[\varphi-(\phi_1+\phi_2)\big]}$$

Define

$$\Delta = \varphi-(\phi_1+\phi_2).$$

Thus,

$$\boxed{I(r) = 2\pi\sum_{m=-\infty}^{\infty} I_m\Big(\frac{2r|\mu_1|}{\sigma_1^2}\Big) I_m\Big(\frac{2|z||\mu_2|}{\sigma_2^2 r}\Big)e^{im\Delta}}$$

### E. Radial Integration

Substitute $I(r)$ back into the expression for $f_z(z)$ in (43):

$$\begin{aligned}
f_z(z) &= \frac{e^{-\frac{|\mu_1|^2}{\sigma_1^2}-\frac{|\mu_2|^2}{\sigma_2^2}}}{\pi^2\sigma_1^2\sigma_2^2}\int_0^{\infty}\exp\Big(-\frac{r^2}{\sigma_1^2}-\frac{|z|^2}{\sigma_2^2 r^2}\Big)\Big\{2\pi \sum_{m=-\infty}^{\infty} e^{im\Delta} I_m\Big(\frac{2r|\mu_1|}{\sigma_1^2}\Big) I_m\Big(\frac{2|z||\mu_2|}{\sigma_2^2 r}\Big)\Big\}\frac{dr}{r} \\
&= \frac{2\,e^{-\frac{|\mu_1|^2}{\sigma_1^2}-\frac{|\mu_2|^2}{\sigma_2^2}}}{\pi\sigma_1^2\sigma_2^2}\sum_{m=-\infty}^{\infty} e^{im\Delta}\,J_m(|z|)
\end{aligned}$$

where

$$J_m(|z|) = \int_0^{\infty}\exp\Big(-\frac{r^2}{\sigma_1^2}-\frac{|z|^2}{\sigma_2^2 r^2}\Big) I_m\Big(\frac{2r|\mu_1|}{\sigma_1^2}\Big) I_m\Big(\frac{2|z||\mu_2|}{\sigma_2^2 r}\Big)\frac{dr}{r}$$

### F. Change of Variable in the Radial Integral

Let $t=r^2$; then $dr=\frac{dt}{2\sqrt{t}}$ and

$$\frac{dr}{r}=\frac{dt}{2t}.$$

Thus,

$$J_m(|z|) = \frac{1}{2}\int_0^{\infty}\exp\Big(-\frac{t}{\sigma_1^2}-\frac{|z|^2}{\sigma_2^2 t}\Big) I_m\Big(\frac{2|\mu_1|}{\sigma_1^2}\sqrt{t}\Big) I_m\Big(\frac{2|z||\mu_2|}{\sigma_2^2}\frac{1}{\sqrt{t}}\Big)\frac{dt}{t}$$

As shown in [22] (see sections around 6.633-6.643), the integral is in a standard form and can be expressed in terms of the confluent hypergeometric function of the second kind $U(a,b,z)$.

$$J_m(|z|) = \frac{1}{2m!}\left(\frac{|\mu_1\mu_2||z|}{\sigma_1^2\sigma_2^2}\right)^m U\Big(m+1,\,1,\,\frac{2|z|}{\sigma_1\sigma_2}\Big)$$

### G. Final Expression for the PDF

Substituting the result for $J_m(|z|)$ into the expression for $f_z(z)$ and combining the contributions from negative and positive $m$ (using $I_{-m}=I_m$ so that the complex exponentials combine into cosines), we get

$$\begin{aligned}
f_z(z) &= \frac{2\,e^{-\frac{|\mu_1|^2}{\sigma_1^2}-\frac{|\mu_2|^2}{\sigma_2^2}}}{\pi\sigma_1^2\sigma_2^2}\left\{\frac{1}{2}J_0(|z|)+\sum_{m=1}^{\infty}\cos(m\Delta)\,J_m(|z|)\right\} \\
&= \frac{2\,e^{-\frac{|\mu_1|^2}{\sigma_1^2}-\frac{|\mu_2|^2}{\sigma_2^2}}}{\pi\sigma_1^2\sigma_2^2}\left\{\frac{1}{2}U\Big(1,1,\frac{2|z|}{\sigma_1\sigma_2}\Big) + \sum_{m=1}^{\infty}\frac{1}{m!}\left(\frac{|\mu_1\mu_2||z|}{\sigma_1^2\sigma_2^2}\right)^m U\Big(m+1,1,\frac{2|z|}{\sigma_1\sigma_2}\Big). \; \cos\Big(m\Big[\varphi-(\phi_1+\phi_2)\Big]\Big)\right\}
\end{aligned} \tag{44}$$

Since the exact final distribution (expressed in terms of confluent hypergeometric functions) is analytically intractable for classifier design, we adopt a moment-matching approximation instead.

### H. Moment Matching

For $x_1$ and $x_2$ we have:

$$\begin{aligned}
\mathbb{E}[x_1]&=\mu_1, \quad \mathbb{E}[|x_1|^2]=\sigma_1^2+|\mu_1|^2, \\
\mathbb{E}[x_2]&=\mu_2, \quad \mathbb{E}[|x_2|^2]=\sigma_2^2+|\mu_2|^2.
\end{aligned}$$

*Step 2: Moments of the Product $z=x_1x_2$*

Because $x_1$ and $x_2$ are independent, the first moment of $z$ is

$$\mu_z=\mathbb{E}[x_1x_2]=\mathbb{E}[x_1]\,\mathbb{E}[x_2]=\mu_1\mu_2.$$

Next, consider the second moment, which is the expected squared magnitude:

$$\begin{aligned}
\mathbb{E}[|z|^2] &= \mathbb{E}[|x_1x_2|^2]=\mathbb{E}[|x_1|^2]\,\mathbb{E}[|x_2|^2] \\
&= (\sigma_1^2+|\mu_1|^2)(\sigma_2^2+|\mu_2|^2) \\
&= \sigma_1^2\,\sigma_2^2+\sigma_1^2\,|\mu_2|^2+\sigma_2^2\,|\mu_1|^2+|\mu_1\,\mu_2|^2
\end{aligned}$$

$$\sigma_z^2 = \mathbb{E}[|z|^2] - |\mu_1\,\mu_2|^2 \\ = \sigma_1^2\,\sigma_2^2 + \sigma_1^2\,|\mu_2|^2 + \sigma_2^2\,|\mu_1|^2 \tag{45}$$

Thus we can now approximate $z$ by a complex Gaussian distribution

$$z \sim \mathcal{CN}(\mu_z, \sigma_z^2) \\ \sim \mathcal{CN}\Big(\mu_1\mu_2,\ \sigma_1^2\,\sigma_2^2 + \sigma_1^2\,|\mu_2|^2 + \sigma_2^2\,|\mu_1|^2\Big) \tag{46}$$

## Appendix B<br>Calculation of PCA using our Model

Consider the distributions:

$$x_0 \sim \mathcal{CN}(\mu_0,\, \sigma_0^2 I_N), \quad x_1 \sim \mathcal{CN}(\mu_1,\, \sigma_1^2 I_N).$$

Assume the two classes occur with equal priors, i.e. $\pi_0 = \pi_1 = \frac{1}{2}$. Our goal is to derive the overall (mixture) covariance of the combined data.

### A. Mixture Model

The overall data distribution is given by the mixture

$$x \sim \frac{1}{2}\,\mathcal{CN}(\mu_0,\, \sigma_0^2 I_N) + \frac{1}{2}\,\mathcal{CN}(\mu_1,\, \sigma_1^2 I_N).$$

*1) Mean:* The mean of the mixture is computed as:

$$\bar{\mu} = \mathbb{E}[x] = \frac{1}{2}\,\mu_0 + \frac{1}{2}\,\mu_1 = \frac{\mu_0 + \mu_1}{2}.$$

*2) Second Moment:* For a complex Gaussian, we have

$$\mathbb{E}[x_i\, x_i^H] = \sigma_i^2 I_N + \mu_i\, \mu_i^H, \quad i = 0, 1.$$

Thus, the second moment of the mixture is

$$\mathbb{E}[x\, x^H] = \frac{1}{2}\,\mathbb{E}[x_0\, x_0^H] + \frac{1}{2}\,\mathbb{E}[x_1\, x_1^H] \\ = \frac{1}{2}\Big(\sigma_0^2 I_N + \mu_0\, \mu_0^H\Big) + \frac{1}{2}\Big(\sigma_1^2 I_N + \mu_1\, \mu_1^H\Big) \\ = \frac{\sigma_0^2 + \sigma_1^2}{2}\, I_N + \frac{1}{2}\Big(\mu_0\, \mu_0^H + \mu_1\, \mu_1^H\Big).$$

*3) Covariance:* The covariance matrix $\Sigma_{\text{mix}}$ is defined by

$$\Sigma_{\text{mix}} = \mathbb{E}[x\, x^H] - \bar{\mu}\, \bar{\mu}^H.$$

$$\bar{\mu}\,\bar{\mu}^H = \left(\frac{\mu_0 + \mu_1}{2}\right)\left(\frac{\mu_0 + \mu_1}{2}\right)^H = \frac{1}{4}\,(\mu_0{+}\mu_1)(\mu_0{+}\mu_1)^H.$$

Expanding the outer product,

$$\bar{\mu}\,\bar{\mu}^H = \frac{1}{4}\Big(\mu_0\, \mu_0^H + \mu_0\, \mu_1^H + \mu_1\, \mu_0^H + \mu_1\, \mu_1^H\Big).$$

Now, substituting into $\Sigma_{\text{mix}}$:

$$\Sigma_{\text{mix}} = \frac{\sigma_0^2 + \sigma_1^2}{2}\, I_N + \frac{1}{2}(\mu_0\, \mu_0^H + \mu_1\, \mu_1^H) \\ - \frac{1}{4}\Big(\mu_0\, \mu_0^H + \mu_0\, \mu_1^H + \mu_1\, \mu_0^H + \mu_1\, \mu_1^H\Big) \\ = \frac{\sigma_0^2 + \sigma_1^2}{2}\, I_N + \frac{1}{4}\Big[2(\mu_0\, \mu_0^H + \mu_1\, \mu_1^H) \\ - (\mu_0\, \mu_0^H + \mu_0\, \mu_1^H + \mu_1\, \mu_0^H + \mu_1\, \mu_1^H)\Big] \\ = \frac{\sigma_0^2 + \sigma_1^2}{2}\, I_N + \frac{1}{4}\Big[\mu_0\, \mu_0^H - \mu_0\, \mu_1^H - \mu_1\, \mu_0^H + \mu_1\, \mu_1^H\Big] \\ = \frac{\sigma_0^2 + \sigma_1^2}{2}\, I_N + \frac{1}{4}\,(\mu_0 - \mu_1)(\mu_0 - \mu_1)^H.$$

Defining

$$\alpha = \frac{\sigma_0^2 + \sigma_1^2}{2}, \quad \beta = \frac{1}{4}, \quad u = \mu_0 - \mu_1.$$

Then the final expression for the mixture covariance is

$$\boxed{\Sigma_{\text{mix}} = \alpha\, I_N + \beta\, u\, u^H} \tag{47}$$

## Appendix C<br>Bhattacharyya Bound (Continuation from Appendix B)

After applying PCA (projection using matrix $U_P \in \mathbb{C}^{N\times P}$ with $U_P^H U_P = I_P$), the data in the $P$-dimensional subspace (with $P \leq N$) becomes

$$z = U_P^H x$$

Thus, for each case the distribution of the projected data is

$$z_0 \sim \mathcal{CN}(m_0,\, \sigma_0^2 I_P), \quad \text{with } m_0 = U_P^H \mu_0 \\ z_1 \sim \mathcal{CN}(m_1,\, \sigma_1^2 I_P), \quad \text{with } m_1 = U_P^H \mu_1$$

### A. Derivation of the Bhattacharyya Distance

For two multivariate complex Gaussian distributions

$$z_i \sim \mathcal{CN}(m_i,\, \Sigma_i) \quad \text{for } i = 0, 1,$$

with

$$\Sigma_0 = \sigma_0^2 I_P, \quad \Sigma_1 = \sigma_1^2 I_P,$$

the Bhattacharyya distance $D_B$ is given by

$$D_B = \frac{1}{8}(m_1 - m_0)^H \Sigma^{-1}(m_1 - m_0) + \frac{1}{2}\ln\frac{\det\Sigma}{\sqrt{\det\Sigma_0\ \det\Sigma_1}}$$

*1) First Term:*

$$\Sigma = \frac{\Sigma_0 + \Sigma_1}{2} = \frac{\sigma_0^2 + \sigma_1^2}{2}\, I_P \\ \Longrightarrow\ \Sigma^{-1} = \frac{2}{\sigma_0^2 + \sigma_1^2}\, I_P$$

the first term becomes

$$\frac{1}{8}(m_1 - m_0)^H \left(\frac{2}{\sigma_0^2 + \sigma_1^2} I_P\right)(m_1 - m_0) \\ = \frac{1}{4(\sigma_0^2 + \sigma_1^2)}\|m_1 - m_0\|^2$$

*2) Second Term:* We have

$$\det \Sigma_0 = (\sigma_0^2)^P, \quad \det \Sigma_1 = (\sigma_1^2)^P$$

and

$$\det \Sigma = \left(\frac{\sigma_0^2 + \sigma_1^2}{2}\right)^P$$

Thus,

$$\frac{1}{2} \ln \frac{\det \Sigma}{\sqrt{\det \Sigma_0 \, \det \Sigma_1}} = \frac{1}{2} \ln \frac{\left(\frac{\sigma_0^2+\sigma_1^2}{2}\right)^P}{(\sigma_0 \sigma_1)^P} = \frac{P}{2} \ln \left(\frac{\sigma_0^2 + \sigma_1^2}{2\sigma_0\sigma_1}\right).$$

### B. Final Expression for the Bhattacharyya Distance

Combining the two terms, we obtain

$$\boxed{\begin{aligned} D_B &= \frac{1}{4(\sigma_0^2 + \sigma_1^2)} \|m_1 - m_0\|^2 + \frac{P}{2} \ln \left(\frac{\sigma_0^2 + \sigma_1^2}{2\sigma_0\sigma_1}\right) \\ &= \frac{1}{4(\sigma_0^2 + \sigma_1^2)} \|U_P^H(\mu_1 - \mu_0)\|^2 + \frac{P}{2} \ln \left(\frac{\sigma_0^2 + \sigma_1^2}{2\sigma_0\sigma_1}\right) \end{aligned}} \tag{48}$$

### C. Classification Error Bound

For equal priors (i.e., $\pi_0 = \pi_1 = \frac{1}{2}$), the probability of error $P_e$ is bounded by

$$P_e \leq \sqrt{\pi_0 \pi_1} \, e^{-D_B} = \frac{1}{2} \, e^{-D_B} \tag{49}$$

### D. Relating the PCA Eigenvalue to the Classification Bound

When we perform PCA on the original data (with a mixture covariance that may have a rank-one update due to the difference in the means), the covariance can be written as

$$\Sigma_{\text{mix}} = \frac{\sigma_0^2 + \sigma_1^2}{2} I_N + \frac{1}{4} (\mu_1 - \mu_0)(\mu_1 - \mu_0)^H.$$

In an idealized scenario where the only nonzero contribution (beyond the isotropic noise) comes from the difference in the means, the dominant eigenvalue of $\Sigma_{\text{mix}}$ is

$$\begin{aligned} \lambda_1 &= \frac{\sigma_0^2 + \sigma_1^2}{2} + \frac{1}{4} \|m_1 - m_0\|^2 \\ &= \frac{\sigma_0^2 + \sigma_1^2}{2} + \frac{1}{4} \|U_P^H(\mu_1 - \mu_0)\|^2 \end{aligned} \tag{50}$$

The term $\frac{1}{4}\|m_1 - m_0\|^2$ is the *additional variance* along the direction given by the difference in the means. Notice that in the Bhattacharyya distance the first term is

$$\frac{1}{4(\sigma_0^2 + \sigma_1^2)} \|m_1 - m_0\|^2 \tag{51}$$

which normalizes the separation of the projected means by the overall noise level. A larger $\|m_1 - m_0\|^2$ increases the dominant eigenvalue, thereby increasing $D_B$.

In other words, we can define the gap between the dominant eigenvalue and the other (noise) eigenvalues the ***Sensing SNR***, and a larger Sensing SNR leads to better discriminability.